\documentclass[twocolumn,trackchanges]{aastex631}
\usepackage{rotating}
\usepackage{tablefootnote}
\usepackage{hyperref}
\usepackage[flushleft]{threeparttable}
\usepackage{tabularx,booktabs}

\let\tablenum\relax
\usepackage{siunitx}
\usepackage{float}
\usepackage{lineno}
\usepackage[caption=false]{subfig}

\newcommand{\hii}{H\thinspace{\sc ii}}
\newcommand{\hi}{H\thinspace{\sc i}}

\newcommand{\fr}{$f_{esc}^{rel}$}
\newcommand{\fa}{$f_{esc}^{abs}$}
\newcommand{\fesco}{$f_{esc}^{abs}$}
\newcommand{\wa}{\si\angstrom}
\newcommand{\Z}{$\rm Z_{\star}$}

\newcommand{\lya}{Ly$\alpha$}

\newcommand{\ewlya}{$W_{Ly\alpha}$}
\newcommand{\ewlyap}{$W_{Ly\alpha}^{prod}$}
\newcommand{\ewlyao}{$W_{Ly\alpha}^{RF}$}
\newcommand{\flyc}{$f_{\rm esc}^{\rm LyC}$}

\newcommand{\tauism}{$\tau_{~ISM}^{~\lambda1216}$}

\newcommand{\flya}{$f_{esc}^{Ly\alpha}$}

\definecolor{mycol}{RGB}{210,25,46}

\shorttitle{LyC $-$ \lya\ relation at $z\sim2.3$}
\shortauthors{A. Citro et al.}

\begin{document}



\title{Challenging the LyC $-$ \lya\ relation: strong \lya\ emitters without LyC leakage at $z\sim2.3$}

\correspondingauthor{Annalisa Citro}
\email{acitro@umn.edu}
 
\author[0009-0000-9676-0538]{Annalisa Citro}
\affiliation{Minnesota Institute for Astrophysics, School of Physics and Astronomy, University of Minnesota, 316 Church Street SE, Minneapolis, MN 55455, USA}

\author[0000-0002-9136-8876]{Claudia M. Scarlata}
\affiliation{Minnesota Institute for Astrophysics, School of Physics and Astronomy, University of Minnesota, 316 Church Street SE, Minneapolis, MN 55455, USA}

\author[0000-0002-6016-300X]{Kameswara B. Mantha}
\affiliation{Minnesota Institute for Astrophysics, School of Physics and Astronomy, University of Minnesota, 316 Church Street SE, Minneapolis, MN 55455, USA}

\author[0000-0002-6039-8706]{Liliya R. Williams}
\affiliation{Minnesota Institute for Astrophysics, School of Physics and Astronomy, University of Minnesota, 316 Church Street SE, Minneapolis, MN 55455, USA}

\author[0000-0002-9946-4731]{Marc Rafelski}
\affiliation{Space Telescope Science Institute, 3700 San Martin Drive, Baltimore, MD 21218, USA}
\affiliation{Department of Physics and Astronomy, Johns Hopkins University, Baltimore, MD 21218, USA}

\author[0000-0002-4917-7873]{Mitchell Revalski}
\affiliation{Space Telescope Science Institute, 3700 San Martin Drive, Baltimore, MD 21218, USA}

\author[0000-0001-8587-218X]{Matthew J. Hayes}
\affiliation{Stockholm University, Department of Astronomy and Oskar Klein Centre for Cosmoparticle Physics, AlbaNova University Centre, SE-10691, Stockholm, Sweden}

\author[0000-0002-6586-4446]{Alaina Henry}
\affiliation{Space Telescope Science Institute, 3700 San Martin Drive, Baltimore, MD 21218, USA}

\author[0000-0001-7016-5220]{Michael J. Rutkowski}
\affiliation{Minnesota State University-Mankato, Telescope Science Institute, TN141, Mankato, MN 56001, USA}

\author[0000-0002-7064-5424]{Harry I. Teplitz}
\affiliation{IPAC, California Institute of Technology, 1low dust content E. California Blvd., Pasadena CA, 91125, USA}

\author[0000-0002-5688-0663]{Andrea Grazian}
\affil{INAF--Osservatorio Astronomico di Padova,
Vicolo dell'Osservatorio 5, I-35122, Padova, Italy}

\author[0000-0002-8630-6435]{Anahita Alavi}
\affiliation{IPAC,  California Institute of Technology, 1low dust content E. California Blvd., Pasadena CA, 91125, USA}

\begin{abstract}

The escape fraction of LyC ionizing radiation (\flyc) is crucial for understanding reionization, yet {difficult} to measure at {$z \gtrsim 4$}. Recently, studies have focused on calibrating indirect indicators of \flyc\ at $z \sim 0.3$, finding that \lya\ is closely linked to it. What is still unclear is whether the LyC $-$ \lya\ relation evolves with redshift, and if \lya\ is truly applicable as an \flyc\ indicator during the reionization epoch. In this study, we investigate seven $-21 \lesssim M_{UV} \lesssim-19$ gravitationally lensed galaxies from the BELLS GALLERY Survey at $z\sim2.3$. {Our targets have rest-frame \lya\ equivalent widths between 40 \wa\ and 200 \wa\ and low dust content ($-2.4 \lesssim \beta \lesssim -2.0$),} both indicative of high LyC escape. {Surprisingly, direct estimates of \flyc\ using Hubble Space Telescope imaging with F275W and F225W reveal that our targets are not LyC emitters, with an absolute \flyc\  $< 13 \%$ (assuming the median IGM transmission)}. The low \flyc, coupled with the high \lya\ equivalent width and escape fraction, could potentially be attributed to the redshift evolution of the neutral hydrogen column density and dust content, {as well as covering fractions of optically thick gas below 1 in high redshift galaxies. Additionally, our analysis suggests that the emission for each lensed component is uniformly absorbed.}
Our results challenge the validity of the extrapolation of $z\sim0$ Ly$\alpha$-based LyC indirect estimators into the reionization epoch. \\

\noindent
\textit{\href{https://astrothesaurus.org/concept-select/}{Unified Astronomy Thesaurus concepts}}: Reionization (1383); High-redshift galaxies (734); Lyman-alpha galaxies (978); Strong gravitational lensing (1643); HST photometry (756)
\end{abstract}

\section{Introduction}
\label{sec:intro}

The epoch of reionization represents the last phase transition of the Universe from neutral to ionized. One of the key parameters necessary to understand this process is the fraction of ionizing radiation (LyC radiation) capable of escaping galaxies' interstellar and circumgalactic medium (ISM and CGM) to eventually ionize the diffuse intergalactic medium {(IGM)}. However, estimating the LyC escape fraction (\flyc) during the epoch of reionization is {very challenging}, due to the near-zero median transmission of the IGM at {redshifts $z\gtrsim4$} \citep[e.g.,][]{YoshiiPeterson1994, Madau1995, Inoue+2014}. Consequently, most studies have focused on utilizing available direct measurements of \flyc\ at low redshifts to calibrate \textit{indirect \flyc\ estimators}, which are based on non-ionizing radiation and are observable at any redshifts. {Once the relations are calibrated at lower redshift, the hope is that they can be applied to galaxies in the epoch of reionization to infer their \flyc.}

In the last decade, the number of direct detections of \flyc\ in nearby galaxies ($z\sim0.3$) has greatly increased, especially due to the success of surveys targeting Green Pea galaxies, combined with the superior sensitivity of the Cosmic Origin Spectrograph on board of the Hubble Space Telescope (HST) \citep[e.g.,][]{Izotov+2016a, Izotov+2016b, Izotov+2018a, Izotov+2018b, Izotov+2021, Malkan+2021, Flury+2022a, Flury+2022b}. LyC detections in local sources have facilitated the calibration of \flyc\ indirect estimators such as the [\ion{O}{3}]$\lambda$5007/[\ion{O}{2}]$\lambda\lambda$3726,9 ratio \citep{JaskotOey2013, NakajimaOuchi2014, Marchi+2018, Mascia+2023}, the \ion{Mg}{2} emission \citep{Henry+2018, Chisholm+2020, Witstok+2021}, the slope of the non-ionizing UV continuum \citep{Zackrisson+2013, Zackrisson+2017, Chisholm+2022}, the equivalent width of Balmer emission lines \citep{Bergvall+2013}, the strength and saturation of the ISM absorption lines \citep{Heckman+2011, Alexandroff+2015, Vasei+2016, Gazagnes+2018, Gazagnes+2020, Saldana-Lopez+2022}, and the star formation surface density \citep{Heckman+2001, Naidu+2020}. In the local Universe, the \lya\ emission appears to be robustly correlated to \flyc, with stronger \lya\ emitters being more efficient LyC leakers. Additionally, using data from the Lyman-$\alpha$ and Continuum Origins Survey (LaCOS) \citep{LeReste+2025}, \citet{SaldanaLopez+2025} have recently found an anti-correlation between LyC emission and extent of the \lya\ halo. {The connection between LyC and \lya\ emission arises because both require an optically thin \hi\ medium to escape the galaxy. 
However, since \lya\ photons can scatter off neutral hydrogen and shift in both frequency and space, the \flyc\ vs. \lya\ correlation is usually characterized by scatter \citep{Flury+2022b}.}

The calibration of indirect estimators, including \lya, has proven significantly more challenging at high redshift ($z\gtrsim 2$). Here, the majority of direct studies have yielded LyC non-detections, with only a handful of confirmed LyC emitters \citep[see ][]{Liu+2023}. Notable {individual} cases are \textit{Ion2} and \textit{Ion3} \citep{Vanzella+2016, Vanzella+2018}, Q1549-C25 \citep{Shapley+2016} A2218-Flanking \citep{Bian+2017}, the Sunburst Arc \citep{Rivera-Thorsen+2019}, and AUDFs01 \citep{Saha+2020}. 

{Several factors affect the ability to obtain direct estimates of LyC emission at high redshift, with one of the most significant being the 
presence of neutral hydrogen in the IGM along the sightlines to LyC emitters. Besides the IGM transmission dropping towards zero at redshifts $z \gtrsim 4$, even at $z \sim 2-3$ ground-based surveys often detect galaxies contaminated by non-ionizing UV emission from low-redshift interlopers \citep{Vanzella+2010}.}

Modeling the IGM transmission also presents its own challenges. Typically, astronomers use a statistical approach, assuming the median IGM transmission among simulated sightlines to the redshift of the LyC candidates. However, due to the bimodal shape of the IGM probability distribution function, the average value does not necessarily correspond to the most likely one \citep{ByrohlGronke2020}, leading to biases in the measurements.  In this regard, \citet{Begley+2022} proposed forward modeling of the flux in the LyC band in order to reduce the bimodality problem.

{The bimodality issue is less significant at $z\lesssim 2.4$, where the stocasticity of the IGM is lower \citep{Lusso+2015}}. Moreover, it can be mitigated when galaxy samples contain only detected LyC galaxies, as the peak of the bimodal distribution centered on zero transmission disappears, making the mean or median of the distribution more sensible and reasonable to use. However, even restricting samples to LyC detected galaxies is not a definitive solution. \citet{Bassett+2021} demonstrated that assuming an median IGM transmission can overestimate the escape fraction in samples of LyC detected galaxies, since their detection implies they are observed through clearer sightlines. Moreover, the sightline-to-sightline variability in IGM opacity makes the interpretations of individual LyC detections difficult. 

{Stacked analyses address this issue by averaging the LyC emission from multiple galaxies, thereby ``smoothing out'' the IGM variability along different lines of sight. Over the past years, stacked analyses have been extensively used to investigate the relationship between LyC and \lya\ at high redshifts, but they have not yielded consistent results. For instance, \citet{Steidel+2018} and \citet{Pahl+2021} found a positive correlation between \flyc\ and the \lya\ equivalent width for galaxies at $z \approx 3$ observed in the Keck Lyman Continuum Spectroscopic Survey (KLCS). Similarly, \citet{Marchi+2018} noted that the flux density ratio of ionizing to non-ionizing photons increases with the \lya\ equivalent width at $z \sim 3.5-4.3$, suggesting that stronger \lya\ emitters are also stronger producers of ionizing photons. \citet{Begley+2022} also reported a positive correlation between \flyc\ and \lya\ photons in VANDELS galaxies at $z \sim 2$. In contrast, \citet{BianFan2020} did not find any individual LyC leaker candidates among their sample of 54 LAEs at $z \approx 3.1$ in the GOODS-S field, nor did they detect any LyC emission in a stack of those objects.  \citet{Fletcher+2019} found a detection rate of 20 \% in a sample of $\sim$ 60 \lya\ emitters at $z \sim 3.1$ from the LymAn Continuum Escape Survey (LACES). However, they found that stacking the \lya\ emitters with no individual LyC detection in their sample yields a stringent upper limit on \flyc\ of $\sim$ 0.5 \%. } 

In addition to the challenges posed by the IGM, high redshift studies face the difficulty of observing faint sources. This issue is particularly problematic given that recent studies are suggesting that low-mass galaxies, with $M_{UV} > -18$, might be the primary contributors to reionization \citep{Lin+2024, Mascia+2024} {which, in turn, would have strong effects on the timescales of reionization \citep{Munoz+2024, Kageura+2025, Umeda+2025}}. While stacked analyses improve the signal-to-noise ratio of individual sources by combining them and help address the IGM variability issue, they have mainly concentrated on magnitudes around $M_{UV} \sim -21$, which is approximately $M^*_{UV}$ at the relevant redshifts \citep[e.g.,][]{Pahl+2024, Marchi+2018, Begley+2022}. Currently, the only other viable way to extend LyC studies to fainter magnitudes at high redshift is gravitational lensing, where the magnification provided by the lens can reveal faint emissions \citep[e.g.,][]{Jung+2024}. However, these studies are still uncommon, with the most notable example being the \textit{Sunburst arc} \citep{Rivera-Thorsen+2019}.



{In this paper, we aim to advance our understanding of the relationship between \flyc\ and \lya\ at high redshift, analyzing a sample of seven $z\sim2.3$ gravitationally lensed galaxies from the BELLS GALLERY Survey \citep{Shu+2016b}. These galaxies are located at the fainter end of the LyC leaker distribution at similar redshifts ($-21 \lesssim M_{UV} \lesssim -19$). We directly measure the LyC emission of these targets and investigate the connection between their \flyc\ and \lya\ emission to determine if \lya\ can serve as an indicator of \flyc\ at higher redshifts.}

The paper is organized as follows: Section \ref{sec:data} details the sample selection (subsection \ref{sec:sample}) and the HST imaging data used for our analysis \ref{sec:hst_data}). Section \ref{sec:reduction} explains how the HST data were reduced (subsection \ref{sec:drizz}) and the Point Spread Functions acquired (subsection \ref{sec:psf}). Section \ref{sec:analysis} covers the data analysis, including the estimation of magnification factors (subsection \ref{sec:mu}), subtraction of the foreground lens from the HST images (subsection \ref{sec:sersic}), extraction of the arc structures from the lensed galaxies (subsection \ref{sec:extr}), measurements of flux and magnitudes (subsection \ref{sec:phot}), determination of the $\beta$ slope and dust extinction of the targets through SED fitting with {\texttt{Bagpipes} }(subsection \ref{sec:dust_ext}), calculation of the LyC escape fraction (subsection \ref{subsec:lyc_emission}), and analysis of the \lya\ emission (subsection \ref{subsec:lya_emission}). Section \ref{sec:results} presents our results and discussion, detailing the relationship between the LyC escape fraction and galaxy properties (subsection \ref{sec:fesc_uvmag}), {the effects of the IGM on the LyC emission (subsection \ref{sec:igm_abs})}, a simple toy model to describe the \lya\ radiative transfer and its comparison to data (subsection \ref{sec:factors}), and the comparison between the LyC and \lya\ escape fractions (subsection \ref{sec:explanation}). Conclusions are drawn in Section \ref{sec:conclusions}. Appendix \ref{appendix:lensing} is dedicated to describing the lensing models used to derive the magnification factors of the lensed galaxies. 
Throughout the paper, we assume a WMAP9 cosmology \citep{Hinshaw+2013}, with $H_{0} =69.3\, \rm km/s/Mpc$ and $\Omega_{M} = 0.287$.


\begin{deluxetable*}{r|cccccc}[ht]
\tabletypesize{\footnotesize}
\tablecolumns{7} 
\tablecaption{\label{tab:hst_img}RA, DEC, and exposure times adopted for each of the targets in our sample.} 
\tablehead{\colhead{Target} & \colhead{RA} & \colhead{DEC}  & \colhead{LyC$~^{a,b}$} &   \colhead{(F438W + F814W)$~^a$} &  \colhead{F606W$~^{a,c}$}  &  \colhead{\# Orbits}}
\startdata
SDSSJ002927 & 00 29 27.3600 & +25 44 2.40 &  12,476 (F275W) &3,000 & 6,165 &  6\\
SDSSJ074249 & 07 42 49.6800 & +33 41 49.20 & 10,954 (F275W) & 2,500 & 2,520 & 5 \\
SDSSJ091859 & 09 18 59.2800 & +51 04 51.60 & 13,899 (F275W) & 3,500 & 2,676 & 6\\
SDSSJ111027 & 11 10 27.1200 & +28 08 38.40 & 16,367 (F275W)& 5,400 & 2,504 & 8 \\
SDSSJ111040 & 11 10 40.3200 & +36 49 22.80 & 10,288 (F275W) & 2,000 & 2,540 &  5 \\
SDSSJ120159 & 12 01 59.0400 & +47 43 22.80 & 13,653 (F225W) & 3,500 & 2,624 & 6 \\
SDSSJ234248 & 23 42 48.7200 & -01 20 31.20 & 13,653 (F225W) & 3,500 & 2,484 & 6 \\
\enddata 
\tablenotetext{$\tiny$ a}{Exposure times in seconds.}
\vspace{-2mm}
\tablenotetext{$\tiny$ b}{The filter indicated in parenthesis is the one used to measure the LyC emission.}
\vspace{-2mm}
\tablenotetext{$\tiny$ c}{Archival data (Program ID \# \href{https://archive.stsci.edu/proposal_search.php?mission=hst&id=14189}{14189}).}
\end{deluxetable*}

\section{data}
\label{sec:data}

\subsection{Selection of the galaxy sample}
\label{sec:sample}
The galaxy sample studied in this work was extracted from the Baryon Oscillation Spectroscopic Survey (BOSS) Emission-Line Lens Survey GALaxy-Ly$\alpha$ EmitteR sYstems (BELLS GALLERY, \citealp{Brownstein+2012}). 
{The BOSS Survey was part of the SDSS-III project, and was designed to map the spatial distribution of luminous red galaxies (LRGs) and quasars. It measured the redshifts of 1.5 million LRGs, obtaining spectra covering the wavelength range 3600 to 10,400 \wa, with a spectral resolution $R\approx1560- 2560$.}
The BELLS GALLERY Survey built on the BOSS Survey, aiming to identify low-mass, dark substructures by capitalizing on the intrinsic compactness of high-redshift lensed \lya\ emitters (LAEs). A subsample of 187 LAE candidates with high-significance \lya\ at $3,600~\wa < \lambda < 4,800 $ \wa\ was isolated from the initial sample. \citet{Shu+2016b} observed 21 of these candidates with HST, confirming that 18 are {strongly graviationally lensed by foreground elliptical galaxies}. From the confirmed sample, we specifically choose the eight targets with magnification $\mu \gtrsim 10$ (as per the values documented in \citealp{Shu+2016b}) and redshifts $2.1 \leq z \leq 2.6$. {In the following, we use the BOSS spectra and HST imaging to determine the LyC and Ly$\alpha$ properties of our galaxies. The BOSS aperture was centered on the foreground elliptical galaxies, as intended by the BELLS Gallery Survey. Consequently, the BOSS spectra primarily capture the light from these elliptical galaxies, but they also include the redshifted Ly$\alpha$ emission from the background sources that falls within the 2'' BOSS aperture. We observe that, although the BOSS fiber fully contains the light from the foreground elliptical galaxies, it barely captures the arc structures of our lensed targets (as shown in Figures~\ref{fig:general1} and \ref{fig:general2}). Consequently, all measurements of Ly$\alpha$ fluxes and equivalent widths reported in the remainder of the paper should be regarded as lower limits.}

\begin{deluxetable}{lcc}[h]
\tabletypesize{\footnotesize}
\tablecolumns{2} 
\tablecaption{\label{tab:275}{Rest-frame wavelength windows captured by the LyC filters F275W and F225W. }}
\tablehead{\colhead{\hspace{-.5cm}Target} & \colhead{LyC Filter} & \colhead{$\Delta_{\lambda} [\wa]$}}
\startdata
SDSSJ002927 &  F275W & $663.50 - 905.75$ \\
SDSSJ074249 & F275W  & $681.28 - 930.01$\\
SDSSJ091859 & F275W  & $673.26 - 919.07$\\
SDSSJ111027 &  F275W  &  $673.26 - 919.07$\\
SDSSJ111040 &  F275W  & $654.03 - 892.81$\\
SDSSJ120159 &  F225W  & $635.78 - 960.27$\\
SDSSJ234248 &  F225W  & $608.56 - 919.16$\\
\enddata 
\end{deluxetable}



\subsection{HST imaging}
\label{sec:hst_data}
The galaxies were observed using the Wide Field Camera~3 on board of HST during Cycle 29 (Program ID \# 16734, PI: Claudia Scarlata). 
We imaged the targets in 46 orbits using  the WFC3 UVIS $+$ F225W (or F275W), F438W, and F814W filters\footnote{The  F225W (or F275W), F438W, and F814W data used in this paper can be found in MAST: \dataset[10.17909/6qm4-vr10]{https://doi.org/10.17909/6qm4-vr10}}. {The F275W and F225W  filters were chosen because they  capture the ionizing radiation at $\lambda<$ 912 \wa\ (above and below $z=2.40$, respectively), allowing a direct measurement of the ionizing radiation. Table \ref{tab:275} summarizes the rest-frame wavelength window probed by the F275W and F225W in our galaxies.} In order to minimize the impact of post-flash and read-out noise, optimize PSF sampling, and enable cleaning of cosmic rays, we used half-orbit exposures for the longest UV integration times, also adopting a large dither pattern. Specifically, for each optical band, we obtained at least three dithered exposures to ensure cosmic ray rejection. The dither pattern was also set to be five times larger than the default box value to help minimize the blotchy pattern observed in similar UV imaging \citep[e.g.,][Figure 15]{Rafelski+2015}.
We used the UVIS2-C1K1C-CTE aperture to center the targets  in the bottom half of chip 2, in quadrant C, in order to minimize the effects of inefficiencies in the Charge Transfer. 

We supplemented our newly acquired data with archival images captured with the F606W filter (Program ID \# \href{https://archive.stsci.edu/proposal_search.php?mission=hst&id=14189}{14189}, PI: A. Bolton). One target, SDSSJ075523, was retrospectively removed from the final sample due to observations of the UV filter (F275W) being conducted at two separate times and unfortunately with two different orientations. {Upon performing data analysis}, it was determined that the signal-to-noise ratio was insufficient to properly align exposures differently oriented. The final sample comprises seven galaxies (see Table \ref{tab:hst_img}).

\section{Data reduction} 
\label{sec:reduction}

The data were downloaded from the MAST archive on February 2023, and were processed using the recently improved WFC3/IR Blob Flats \citep{OlszewskiMack2021}. We used the most recent bad-pixel masks available at that time (CRDS context: \href{https://hst-crds.stsci.edu/context_table/hst_1063.pmap}{$\rm hst\_1063.pmap$}). The 128 individual exposures were cleaned for hot pixels and readout cosmic rays (ROCRs), and subsequently aligned and drizzled\footnote{We applied the same method used for the newly acquired filters to process the F606W images.}.  

We cleaned the HST images using the darks HST/WFC pipeline described in \citet{Prichard+2022}. Compared to the standard dark HST/WFC3 pipeline, the new pipeline includes the following improvements: 

 \begin{enumerate} 
    \item A new hot pixel flagging algorithm which uses a variable (instead of the standard constant) threshold as a function of the distance from the readout amplifier. This usually results in $\sim30$ \% more hot pixels identified on average.
    \item A new algorithm that flags the divots corresponding to ROCRs. {These cosmic rays (CRs)} are usually overcorrected by the code that perform the  {charge transfer efficiency (CTE)} correction, because they strike the array while it is being read out.
    \item A new algorithm to equalize the signal levels in the four different amplifier regions of each exposure.
\end{enumerate}

In addition to these three cleaning steps, we corrected the exposures for cosmic rays using our publicly available code\footnote{\url{https://github.com/mrevalski/hst_wfc3_lacosmic}},  which is built upon the \texttt{LACOSMIC} code \citep{VanDokkum2001,McCully+2018}. 

\subsection{Drizzling}
\label{sec:drizz}



The drizzling procedure consists of mapping the pixels of the original exposures into a subsampled output grid that is corrected for geometric distortion. The pixels in the output image are usually shrunk to a fraction of their original size in order to avoid undersampling of the final point spread function (PSF). 

The drizzling process necessitates precise alignment of the exposures, which we achieved using the \textsc{tweakreg} routine from the \textsc{drizzlepac} (v3.5.0) software package \citep{Gonzaga+2012, Hoffmann+2021}.

To initiate the alignment, we employ the \textsc{UPDATEWCS} function with the \textsc{USE\_DB} option set to False. This action eliminates the default WCS coordinates responsible for the misalignment between direct images and grism exposures, restoring the most recent HST guide star-based WCS coordinates. A catalog of GAIA EDR3 sources \citep{GaiaColl2022} with position uncertainties of $<10$ mas is then generated for the alignment.

Initially, the F814W optical image is aligned with GAIA sources. We use default \textsc{tweakreg} parameters, with an average setting of the \textsc{imagefindcfg} threshold at $2\sigma$. This threshold typically yields about 5 suitable sources for alignment. Subsequently, we align the F438W, the UV F275W (F225W) and the archival F606W images to the previously GAIA-aligned F814W image.

After aligning all exposures at the filter level, we generate final mosaics utilizing the \textsc{astrodrizzle} routine from \textsc{drizzlepac} (v3.5.0). With an average of fewer than $\sim10$ dithered exposures for each filter, we decrease the pixel size to 80 \% of the original.

For the F275W and F225W exposures (our LyC images), an additional step is necessary, as they were captured during two distinct visits. Initially, the \textsc{astrodrizzle} routine was utilized to generate visit-level mosaics. Following this, one of the two images is selected as the reference, and the other is aligned to it. Ultimately, we generated a composite drizzle incorporating both visits.

\begin{deluxetable}{ll}[h]
\tabletypesize{\footnotesize}
\tablecolumns{2} 
\tablecaption{\label{tab:astrodrizzle}Main \textsc{astrodrizzle} parameters used to produce the drizzles.} 
\tablehead{\colhead{\hspace{-.5cm}Parameter} & \colhead{\hspace{-1.8cm}Value}}
\startdata
final\_scale     & 0.04 arcsec/pix            \\
final\_pixfrac~  & 0.8              \\
skymethod        & globalmin+match  \\
final\_wht\_type & IVM              \\
combine\_type    & minmed$~^a$ \\   
\enddata 
\tablenotetext{$\tiny$ a}{We use 'minmed' when we have $\leq6$ exposures to combine, and 'imedian' when we have $>6$ exposures to combine.}
\end{deluxetable}

To get an estimate of the depth of the drizzled LyC images, we derive the flux limit for point sources. To do so, we place $N = 1000$ circular apertures with a diameter of $0.2''$ in random source-free regions of the image, and compute the standard deviation of the aperture flux distributions. For the UV filters sensitive to the LyC radiation, we find $1\sigma$ depth $\rm m_{Lim}\sim 29$ (See Table \ref{tab:photometry}). These highly sensitive images enable a precise evaluation of the potential presence of LyC emission. 

\subsection{Acquiring PSFs}
\label{sec:psf}


The modeling of the lens surface brightness profile and the measurement of colors require reliable estimates of the PSF in the various filters.  To obtain these PSFs we adopt the publicly-available code \textsc{hst\_wfc3\_psf\_modeling}\footnote{\url{https://github.com/mrevalski/hst_wfc3_psf_modeling}} from \cite{mitchell_revalski_2022_7458566}.
This code uses two different approaches depending on whether or not there are sufficient isolated, non-saturated stars in the image. In the first case, the code derives the image PSF by stacking all the individual stars together. Instead, if less than two stars are found, the code uses the \textsc{make\_model\_star\_image} tool from the \textsc{psf\_tools.PSFutil} module in the \textsc{wfc3\_photometry}\footnote{\url{https://github.com/spacetelescope/wfc3_photometry}} public package (developed by V. Bajaj). 
This tool plants empirical PSFs (provided on the \href{https://www.stsci.edu/hst/instrumentation/wfc3/data-analysis/psf}{HST Data Analysis} page) into copies of the original input drizzles, which can then be stacked as real stars to produce a PSF. Based on the number of stars found by the code, we use the first method for the F814W images, and the second one for F606W, F438W, and F275W (F225W). Table \ref{tab:img} lists the Full Width Half Maximum (FWHM) of the PSFs in each filter. 

\begin{deluxetable}{rc}[h]
\tabletypesize{\footnotesize}
\tablecolumns{5} 
\tablecaption{\label{tab:img}Full Width at Half Maximum of the PSF of the five filters considered in this work.} 
\tablehead{\colhead{Filter} & \colhead{FWHM (arcsec)}}
\startdata
F275W & 0.087\\
F225W & 0.087\\
F438W & 0.099\\
F606W & 0.094\\
F814W & 0.096\\
\enddata 
\end{deluxetable}

\begin{figure*}[ht]
\centering
\includegraphics[width=0.99\textwidth, trim={0em 8em 0em 0em}]{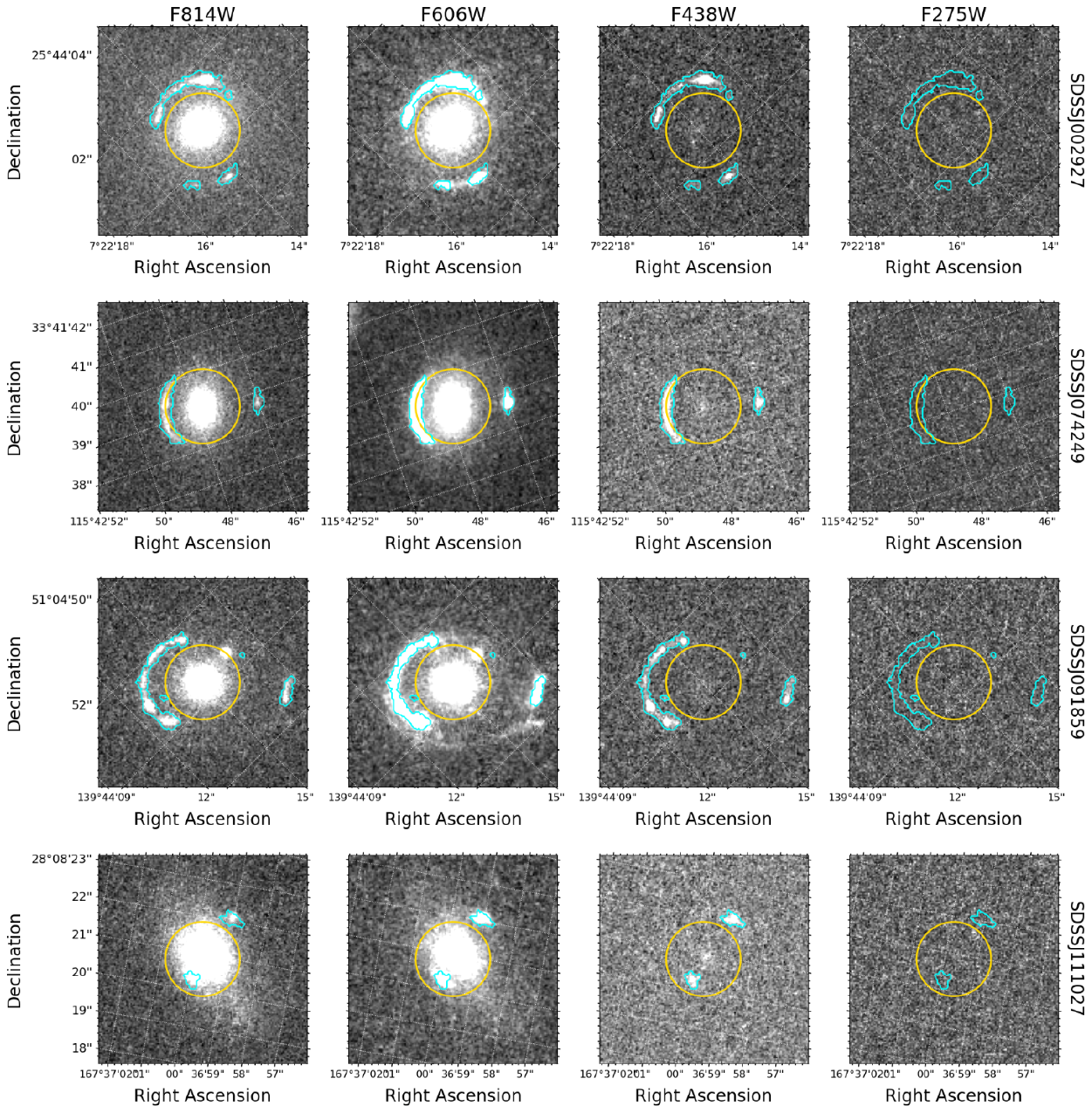}
\caption{Lensed galaxies and extracted arc features. From left to right, we show the F814W, F606W, F438W, and F275W filter images for four of out targets. The cyan regions are the source arc structures extracted from the F438W mosaic using the \textsc{residual\_features\_extraction} code.  The yellow circles show the 1 arcsec radius BOSS fiber. The mosaics shown in the Figure are the drizzled output images from \textsc{astrodrizzle}. To infer the photometry, we subtract the foreground lens galaxy from the mosaics first.}
\label{fig:general1}
\end{figure*}

\begin{figure*}[ht]
\includegraphics[width=0.99\textwidth]{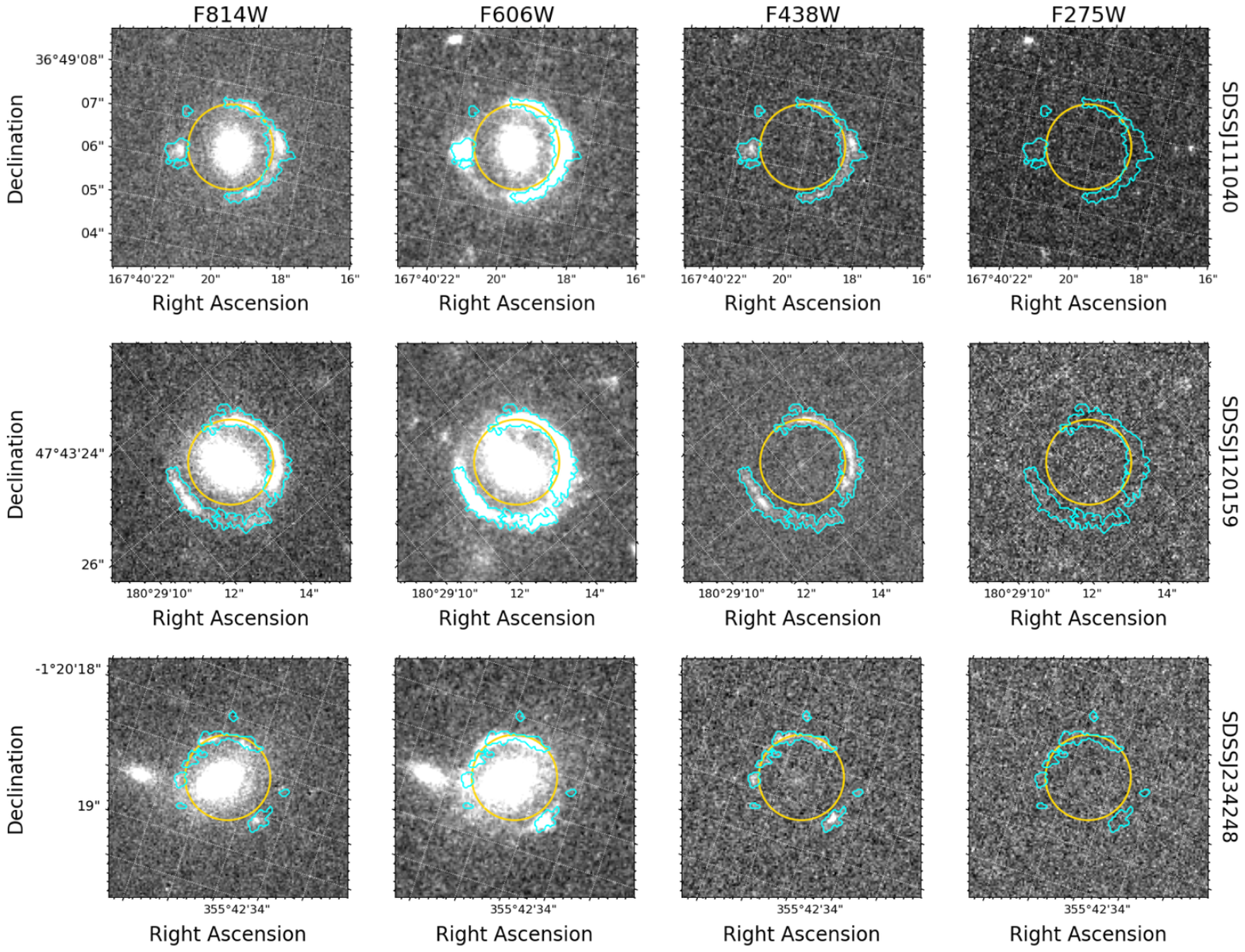}
\caption{Same as Figure \ref{fig:general1} for the three remaining targets of our sample. From left to right, we show the F814W, F606W, F438W, and the LyC filter (F275W  for SDSSJ111040, and F225W for SDSSJ120159 and SDSSJ23248).}
\label{fig:general2}
\end{figure*}

\section{Data analysis}
\label{sec:analysis}

\subsection{Estimation of the magnification factors}
\label{sec:mu}
The first step of our analysis involves quantifying the magnification factors $\mu$ of our gravitationally lensed targets. We carry out our own lensing model to obtain the magnification factor $\mu$ for each one of the seven galaxies. Section \ref{appendix:lensing} of the Appendix describe our modeling in details, while Table \ref{tab:properties} lists the derived magnification factors. In the following, we adopt the more conservative values $\mu_{\rm a}$.

\begin{deluxetable}{r|cccccc}[h]
\tabletypesize{\footnotesize}
\tablecolumns{2} 
\tablecaption{\label{tab:properties} Redshifts and magnification factors of the seven BELLS gallery analyzed in this work.} 
\tablehead{\colhead{Target} & \colhead{$z_{lens}~^a$} & \colhead{$z_{Ly\alpha}~^b$}   & \colhead{$z_{\rm sys}~^c$} & \colhead{$\mu_{\rm S}~^d$}  & \colhead{$\mu_{\rm a}~^e$}  & \colhead{$\mu_{\rm b}~^f$}}
\startdata
SDSSJ002927 &  0.5869 &  2.4504 & 2.4500 &  14 & 12 $\pm$ 5 & 19 $\pm$ 7 \\
SDSSJ074249 & 0.4936 & 2.3633 & 2.3625  & 16 & 17 $\pm$ 5 & 22 $\pm$ 7 \\
SDSSJ091859 & 0.5811 & 2.4030 & 2.4000  & 18 & 11 $\pm$ 4& 13 $\pm$ 5  \\
SDSSJ111027 &  0.6073 & 2.3999 & - & 8 & 5 $\pm$ 3 & 7 $\pm$ 4 \\
SDSSJ111040 &  0.7330 & 2.5024 &- & 17 & 11 $\pm$ 3 & 13 $\pm$ 3 \\
SDSSJ120159 &  0.5628 & 2.1258 & -& 12 & 8 $\pm$ 3 & 14 $\pm$ 8 \\
SDSSJ234248 &  0.5270 & 2.2649 & -& 23 & 7 $\pm$ 3 & 11 $\pm$ 4 \\
\enddata 
\tablenotetext{$\tiny$ a}{Redshift of the foreground lens.}
\vspace{-2mm}
\tablenotetext{$\tiny$ b}{Redshift of the source from \lya\ line.}
\vspace{-2mm}
\tablenotetext{$\tiny$ c}{Redshift of the sources from UV absorption lines 
\citep{MarquesChaves+2020}, or from strong nebular lines (this work); see subsection \ref{subsec:lya_emission} for further details.}
\vspace{-2mm}
\tablenotetext{$\tiny$ d}{Magnification factor from \citet{Shu+2016a}.}
\vspace{-2mm}
\tablenotetext{$\tiny$ e}{Magnification factor calculated considering 0.018 counts/s as a threshold (see Section \ref{appendix:lensing} for further details). }
\vspace{-2mm}
\tablenotetext{$\tiny$ f}{Magnification factor calculated considering 0.025 counts/s as a threshold (see Section \ref{appendix:lensing} for further details).}
\end{deluxetable}

\subsection{S\'ersic Fitting of the lens foreground ellipticals}
\label{sec:sersic}
The second step of the analysis involves quantifying the flux of the lensed galaxies in the observed filters F275W (F225W), F438W, F606W, and F814W. To accomplish this task, we need to eliminate any low-surface brightness flux emitted by the elliptical lens that might be contaminating the light from the background source. Therefore, we fit each foreground lens (in various wavelength bands) with a PSF-convolved 2D S\'ersic model using the \textsc{GALFIT} software \citep{Peng+2002}, and then we subtract its contribution from our images.
To derive the S\'ersic parameters of the foreground lens galaxies, we follow standard procedures employed by many previous studies \citep[e.g.,][]{VanDerWel+2014}. For each galaxy, we start with the F814W image, since it has the highest S/N ratio of all the filters. We perform source extraction of this image using \textsc{SEP} \citep{Barbary+2016} to estimate the gross properties of the lens galaxy, such as apparent magnitude, semi-minor and major axes lengths, and positional angle.
{The segmentation map of the elliptical galaxy is robust, effectively separating it from the surrounding arc structures. This is because, for six out of seven sources, the lensed images are sufficiently distant from the elliptical galaxy. For the one source, SDSSJ111027, we acknowledge that a portion of the lensed source is very close to (and within the envelope of) the elliptical galaxy. However, since the elliptical galaxy’s profile fitting is dominated by its relative brightness, the presence of an overlapping source does not affect the overall fitting process of the elliptical profile.}
By masking all the sources identified in this step, we compute the ``background'' contribution as the mean of the pixel-value distribution in the source masked image. Using these values as initial guesses, we fit the lens galaxy letting all the seven parameters of the S\'ersic model to vary freely, except for the background value, which is left fixed (the background is estimated as described in the following subsection \ref{sec:phot}).

To self-consistently obtain S\'ersic models across the remaining wavelength bands, we rescale the best-fit, PSF-convolved F814W S\'ersic model's magnitude to the magnitudes of the F606W, F438W, and F275W (F225W) filters. These magnitudes are obtained by integrating the flux over a circular aperture of radius $\rm r = 5\times$ the FWHM of the PSF within those bands. Finally, for each filter, we create residual images by subtracting the best-fit S\'ersic model  from the original images. We use these residual images to extract the arc structures of the lensed galaxies and quantify their properties, as explained in the next subsection \ref{sec:extr}.

    

\subsection{Extraction of the Arcs}
\label{sec:extr}
The second step of our analysis consists in establishing the aperture that encompasses the arc structures of our lensed targets. In order to do this, we employ the publicly available residual feature extraction software\footnote{\url{https://github.com/AgentM-GEG/residual_feature_extraction}} \citep{Mantha+2019}. This pipeline is designed to work directly with the outputs of {\tt GALFIT}, and  works towards extracting both flux and area-wise significant contiguous regions of interest within the residual images. This software has proven to be efficient in extracting extended, low-surface brightness signatures within parametric model-subtracted images and has been demonstrated on HST imaging, notably also extracting gravitational arc structures (see Figure 10 in \citealp{Mantha+2019}). One of the primary outputs of the feature extraction method is a binary segmentation map, where the extracted pixels that meet the criteria are labelled as 1 and all else as zero. These segmentation maps, corresponding to the extracted arc structures, serve as custom apertures for downstream photometry estimations.

We apply the \citet{Mantha+2019} method to the residual images in the F438W filter. Since the foreground lens is intrinsically fainter in this band, with this choice we are minimizing any potential residual contribution (if any) from the lens. Once the segmentation map is derived for the F438W residual image, we apply it to the other bands. In Figures\,\ref{fig:general1} and \ref{fig:general2}, we illustrate the extracted arc features (cyan contours) overlaid on top of the multiple wavelength images.

\begin{deluxetable*}{lccccccc}
\tabletypesize{\footnotesize}
\tablecolumns{8} 
\tablecaption{\label{tab:photometry} Photometry of the lensed galaxies analyzed in this work.} 
\tablehead{\colhead{Target} & \colhead{z} & \colhead{$\mu_{a}$} & \colhead{Area$~^a$}  & \colhead{Filter} & \colhead{ $\rm m_{AB}~^b$} & \colhead{$\rm M_{AB}~^c$} & \colhead{m$_{\rm Lim}$} }
\startdata
\hline
       &          &       &     & F814W  &          22.43  $\pm$ 0.03  & -20.06$^{\pm 0.03}_{\pm0.50}$ &  27.17\\
       &          &       &     & F606W  &          22.28 $\pm$ 0.01    & -20.21$^{\pm 0.01}_{\pm0.50}$ & 28.38\\
SDSSJ002927 &    2.45   & 12 $\pm$ 5 & 0.14     & F438W  &     22.19 $\pm$ 0.02 &    -20.30$^{\pm 0.02}_{\pm0.50}$ &  27.13  \\
       &          &       & & F275W  &$\gtrsim$ 28.44    & - &  28.44 \\
\hline
&          &       & & F814W  &           22.41 $\pm$ 0.03   & -19.63$^{\pm0.03}_{\pm0.32}$ &  27.17 \\
        &          &      & & F606W  &           22.35 $\pm$ 0.01     & -19.69$^{\pm0.01}_{\pm0.32}$ & 28.64 \\ 
SDSSJ074249   & 2.36 & 17 $\pm$ 5 & 0.07 & F438W  &         22.18 $\pm$ 0.03  & -19.86$^{\pm0.03}_{\pm0.32}$ &  26.95 \\
       &          &       & & F275W  &  $\gtrsim$ 28.20    & -  &  28.20\\ 
\hline
&          &       & & F814W  &        21.77 $\pm$ 0.02  & -20.78$^{\pm0.02}_{\pm0.40}$ & 27.17\\
       &          &       & & F606W  &     21.49 $\pm$ 0.01     & -21.06$^{\pm0.01}_{\pm0.40}$ &  28.57 \\ 
SDSSJ091859  & 2.40  & 11 $\pm$ 4 & 0.12 & F438W  &       21.31 $\pm$ 0.02 & -21.24$^{\pm0.02}_{\pm0.40}$ &  26.95\\
&          &       & & F275W  &   $\gtrsim$ 28.30   & -  &  28.30 \\ 
\hline
 &          &       & & F814W  &  24.33 $\pm$ 0.04 & -19.07$^{\pm0.04}_{\pm0.70}$ & 28.49\\
        &          &      & & F606W  &   24.27 $\pm$ 0.02 & -19.14$^{\pm0.02}_{\pm0.70}$  & 29.27  \\
SDSSJ111027   & 2.40 & 5 $\pm$ 3 & 0.07  & F438W  &    24.05 $\pm$ 0.04  &  -19.36$^{\pm0.04}_{\pm0.70}$ &  28.38\\
       &          &       & & F275W  &   $\gtrsim$ 29.45    & - &  29.46 \\
\hline
&          &       &  & F814W  &           22.31 $\pm$ 0.02  & -20.31$^{\pm0.02}_{\pm0.30}$ & 26.80\\
         &          &     & & F606W  &         22.20 $\pm$ 0.01   & -20.42$^{\pm0.01}_{\pm0.30}$ & 28.05 \\ 
SDSSJ111040 & 2.50  & 11 $\pm$ 3 & 0.12 & F438W  &           22.24 $\pm$ 0.02  & -20.39$^{\pm0.02}_{\pm0.30}$ & 26.73 \\
       &          &       & & F275W  &  $\gtrsim$ 28.25   & - & 28.25\\ 
     \hline
 &          &             & & F814W     &  21.65 $\pm$ 0.04     & -21.02$^{\pm0.04}_{\pm0.40}$ & 25.98 \\
       &          &             & & F606W     &   21.53 $\pm$ 0.03     & -21.14$^{\pm0.03}_{\pm0.40}$ & 27.25  \\
SDSSJ120159      & 2.13 & 8 $\pm$ 3  & 0.75 & F438W     &   21.39 $\pm$ 0.03     & -21.27$^{\pm0.03}_{\pm0.40}$ &  26.28\\
       &          &            & & F225W     &  $\gtrsim$ 27.83       & - & 27.83 \\
       \hline
       &          &       & & F814W  &  23.70 $\pm$ 0.05   & -19.23$^{\pm0.05}_{\pm0.50}$  &  27.24\\
       &          &       & & F606W  & 23.53 $\pm$ 0.01   & -19.40$^{\pm0.01}_{\pm0.50}$ &  28.62\\
SDSSJ234248 & 2.27   & 7 $\pm$ 3 & 0.12 & F438W  &  23.17  $\pm$ 0.03  & -19.76$^{\pm0.03}_{\pm0.50}$ & 27.27 \\
       &          &      & & F225W  &  $\gtrsim$ 28.45      & - &  28.46 \\
\enddata 
\tablenotetext{$\tiny$ a}{Area of the arc structures in the source plane in arcsec$^2$.}
\vspace{-2mm}
\tablenotetext{$\tiny$ b}{Observed apparent magnitudes in AB system (not corrected for the magnification factor).}
\vspace{-2mm}
\tablenotetext{$\tiny$ c}{ Observed absolute magnitude in each filter, corrected for the magnification factor. The upper error is the error on the photometry, while the lower error is the error on the lensing.}

\end{deluxetable*}

\begin{deluxetable}{r|cc}[h]
\tabletypesize{\footnotesize}
\tablecolumns{2} 
\tablecaption{\label{tab:betaslope} {$\beta$ slopes and color excesses $E(B-V)_{\rm fit}$ for our targets.}}
\tablehead{\colhead{Target} & \colhead{$\beta$ slope $^a$} & \colhead{$E(B-V)_{\rm fit}~^b$}}
\startdata
SDSSJ002927 & -2.35 $\pm$ 0.05 &  0.045\\ 
SDSSJ074249 & -2.00 $\pm$ 0.04  &  0.050 \\
SDSSJ091859 &  -2.28 $\pm$ 0.03  & 0.094\\
SDSSJ111027 & -2.05 $\pm$ 0.10  &  0.045\\
SDSSJ111040 & -2.21 $\pm$ 0.06 & 0.041 \\
SDSSJ120159 & -2.02 $\pm$ 0.07  & 0.052\\
SDSSJ234248 &  -2.42 $\pm$ 0.13 &  0.063\\ 
\enddata 
\tablenotetext{$\tiny$ a}{ Obtained from the $\log(f_{\lambda})$ vs. $\log(\lambda)$ relation between F814W and F606W fluxes.}
\vspace{-2mm}
\tablenotetext{$\tiny$ b}{{Obtained from the SED fitting with \texttt{Bagpipes} and a \citet{Calzetti+2000} attenuation curve.}}
\end{deluxetable}

\subsection{Flux measurements}
\label{sec:phot}

The fluxes and magnitudes of the lensed targets in the various bands are derived by summing the pixel fluxes within the segmentation map of the arc structures obtained in subsection \ref{sec:extr}. We observe that the light captured by our apertures in the LyC images (F225W or F275W) does not display emission above the noise threshold (see Figures~\ref{fig:general1} and \ref{fig:general2}). Nevertheless, it is possible that the presence of small, leaking knots could still hold significance if observed through a smaller aperture targeting a specific clump. Upon thorough examination of our images to investigate this possibility, we have conclusively ruled out the existence of any such small knots.

In order to derive upper limits on the LyC emission, we estimate the background level by dropping $1000$ random, identical circular apertures onto the original F275W and F225W filters images and measuring the background flux in each of them. The apertures are placed exclusively in regions that are free from contamination by background sources, and have the same area as the total number of pixels within the segmentation maps defined from the F438W residual images. The $1\sigma$ upper limits on the LyC emission are defined as the standard deviation of the background flux distribution.
Table \ref{tab:photometry} summarizes the photometry of our targets in the available bands. Note that our magnitudes are corrected for Galactic extinction using the \citet{Schlegel+1998} dust map and a \citet{Cardelli+1989} extinction curve with $R_{V}=2.1$. 


\begin{figure}[H]
\includegraphics[width =\columnwidth]{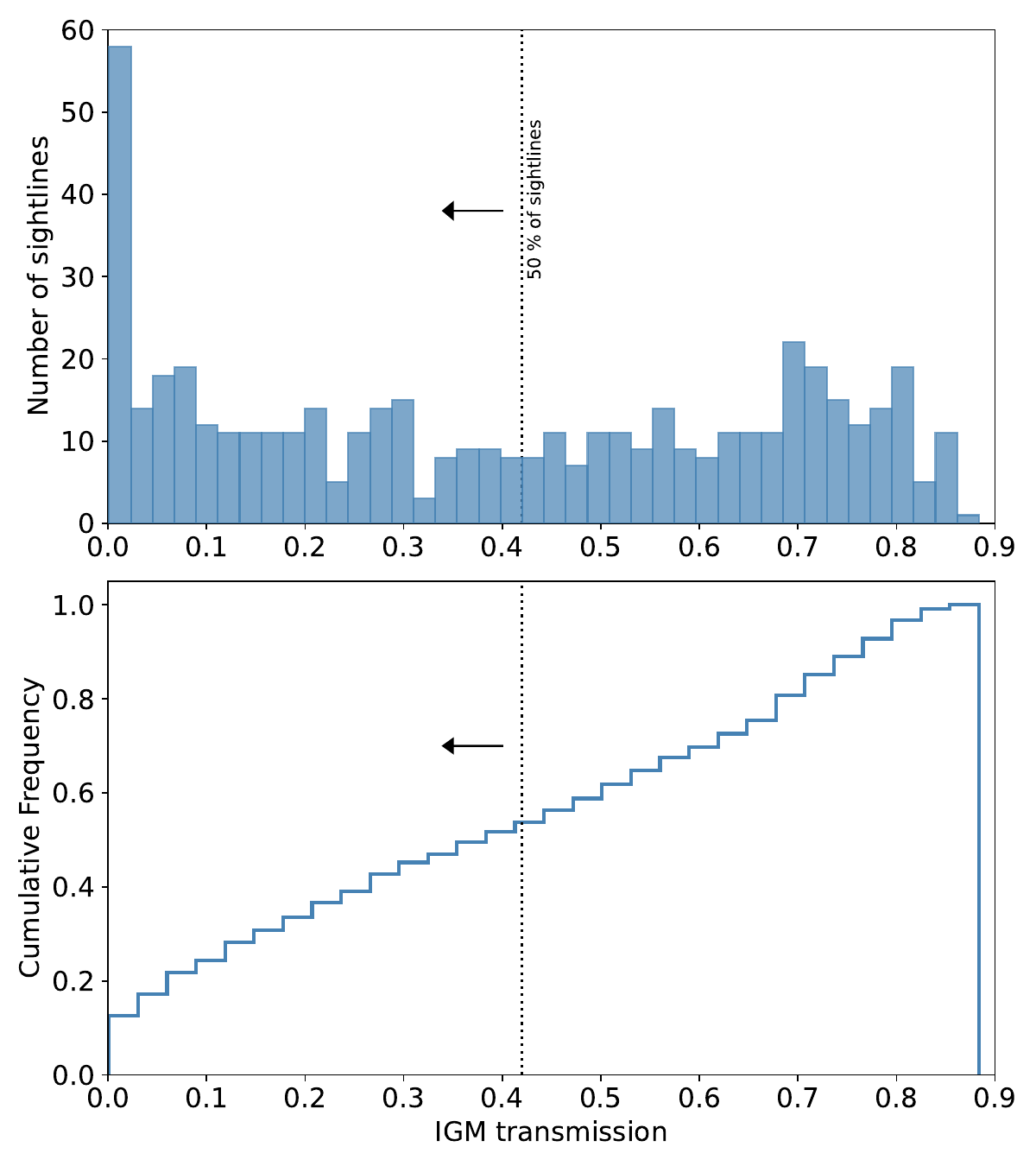}

\caption{{Top panel: histogram of the F275W filter transmission-weighted IGM transmission of the 500 simulated lines of sight for target SDSSJ091859. Bottom panel: cumulative fraction up to any given IGM transmission. The vertical black dotted line shows the 50th percentile of the distribution, indicating that 50 \% of the sightlines have IGM transmission below $e^{-\tau_{\rm IGM}}$ =0.42.}}
\label{fig:igm_tran}
\end{figure}

\subsection{SED fitting and \texorpdfstring{$\beta$}{B} slope}
\label{sec:dust_ext}

We model the galaxies' spectral energy distribution using the Bayesian Analysis of Galaxies for Physical Inference and Parameter Estimation code (\texttt{Bagpipes}, \citealp{Carnall+2018}), in order to derive the   properties needed for the calculation of the LyC escape fractions (e.g., dust attenuation and intrinsic ratio between ionizing and non-ionizing continuum). 
Given the small number of photometric bands, we limited the number of free parameters. Specifically, we modeled the galaxies' SEDs assuming a constant star-formation rate and stellar metallicity of 1/2 solar  (consistent with values observed at the targets' redshifts). Dust attenuation  was allowed to vary, and we assumed both a \citet{Calzetti+2000} and an SMC \citep{Gordon+1998} attenuation curve. \texttt{Bagpipes} uses the 2016 version of the \citet{BruzualCharlot2003} models, and assume a \citet{Chabrier2003} initial mass function. For the identification of the best fit parameters, we used the pure Python nautilus nested sampling algorithm.

Additionally, we use the observed photometry in the F606W and F814W filters to calculate the UV continuum slope $\beta$ of our targets. The error on the slopes is determined as the standard deviation of 100 measurements of $\beta$, obtained by varying the observed magnitudes within their errors. The results of the SED modeling and the derived $\beta$ values are listed in Table \ref{tab:betaslope}.

\subsection{Lyman Continuum escape fraction}
\label{subsec:lyc_emission}
We use the following equation to derive  upper limits on the relative escape fraction  in our targets:

\begin{equation}
\label{eq:frel}
f_{esc}^{rel} = \frac{ L(1500)/L(800)_{\rm int} }{ f(1500)/f(800)_{\rm lim}}\times e^{\tau_{\rm IGM}}~~,
\end{equation}

where $f(1500)/f(800)_{\rm lim}$ is the observed ratio between non-ionizing and ionizing radiation, measured from the F606W and F275W (F225W) fluxes.

{We derive the intrinsic ratio $L(1500)/L(800)_{\rm int}$ from the best fit models described above. Specifically, for each galaxy, we first convolve the best fit models with the F275W (or F225W) and the F606W filter response curves, and then perform the ratio. {Interesingly,  we find 
that the LyC flux derived with the convolution method described above is $\sim$ 10\% lower than the LyC flux derived as average flux within the 880-910 \wa\ wavelength window (which is usually adopted in spectroscopic measurements of LyC, including the LzLCS)}.
${\tau_{\rm IGM}}$ represents the optical depth of LyC photons through the IGM along the line of sight to each individual galaxy. We need to correct for the IGM absorption in order to obtain the true \fr, i.e. the escape fraction in the vicinity of the galaxy right after escaping the ISM. Since the IGM is highly variable, we estimate its median transmission by running a Monte Carlo simulation using known distributions of \ion{H}{1} absorber column densities (following the prescription by \citet{Steidel+2018}). We simulate 500 realizations of the line of sight of each target at their individual redshifts. Then, for each target, we determine the median transmission as the median of the 500 IGM transmissions as a function of wavelength convolved with the F275W (or F225W) filter response curves. Figure \ref{fig:igm_tran} shows the results of our simulations for one of our targets.}

\begin{figure}[h]
\includegraphics[width =\columnwidth]{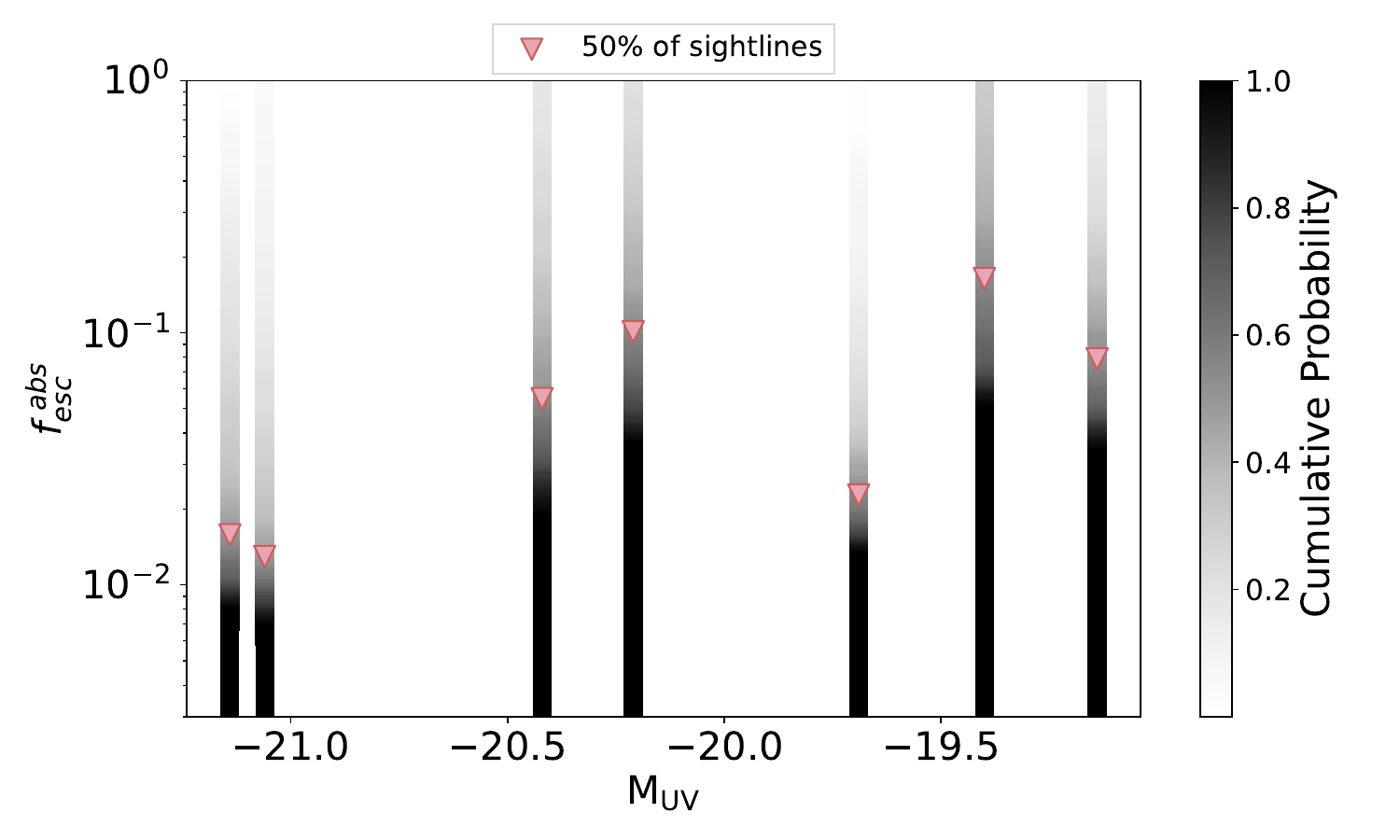}
\caption{{Limits on the absolute escape fraction \fa\ accounting for the IGM transmission as a function of the absolute UV magnitude. The cumulative probability increases from white to black, indicating that more than 50 \% of the sightlines provide an \fa\ $< 13$ \%.}}
\label{fig:fesc_igm}
\end{figure}

From \fr, we derive the absolute escape fractions as \fa\ = \fr\ $\times 10^{-0.4\, A_{1500}}$, where $A_{1500}$ is the dust attenuation at 1,500 \AA. We compute $A_{1500}$  from $E(B-V)_{\rm fit}$ for the two attenuation curves SMC \citep{Gordon+1998}, and \citet{Calzetti+2000}. To maintain consistency with most literature, we present the results obtained from the \citet{Calzetti+2000} curve. This choice aligns with the majority of high-redshift \fa\ measurements and with the local  LzLCS \fa, which use a \citet{Reddy+2016} curve (similar to the Calzetti curve). 
The intrinsic $L(1500)/L(800)_{\rm int}$ and the upper limits on \fr\ and \fa\ are listed in Table \ref{tab:frel}. Moreover, our resulting limits on \fa, accounting for the IGM transmission, are shown in Figure \ref{fig:fesc_igm} as a function of the absolute UV magnitude of our targets. {Throughout the paper, we will report the escape fractions relative to the median IGM transmission, i.e., corresponding to 50 \% of the sightlines.}

\begin{deluxetable}{r|cccc}
\tabletypesize{\footnotesize}
\tablecolumns{5} 
\tablecaption{{\label{tab:frel} Intrinsic ratio between ionizing and non ionizing emission; 1$\sigma$ upper limits on the relative and absolute escape fractions derived from the F275W(F225W) and the F606W data; 50th percentile of the IGM transmission distribution from 500 simulated sightlines. }}
\tablehead{\colhead{Target} & \colhead{$\mathit{\frac{L(\rm 1500)}{L(\rm 800)}}_{\rm int}^{~a}$} & \colhead{$f_{esc}^{rel}~^{b}$} & \colhead{$f_{esc}^{abs}~^{c}$} & \colhead{$e^{-\rm \tau_{IGM}}$}}
\startdata
SDSSJ002927 & \textcolor{black}{5.4} &  0.105 &  0.069 &  0.28 \\ 
SDSSJ074249 & \textcolor{black}{5.9} & 0.028 &  0.018  & 0.46 \\
SDSSJ091859 & \textcolor{black}{5.2} & 0.048   &  0.020 & 0.42\\
SDSSJ111027 & \textcolor{black}{5.3} & 0.202  & 0.132 & 0.34\\
SDSSJ111040 & \textcolor{black}{5.5} & 0.160  &  0.108 & 0.26\\
SDSSJ120159 & \textcolor{black}{5.3} & 0.081  &  0.050 & 0.41\\
SDSSJ234248 & \textcolor{black}{5.7} & 0.210  &  0.115 & 0.24\\
\enddata 
\tablenotetext{$\tiny$ a}{Intrinsic ratio corrected for dust extinction using $E(B-V)_{fit}$ and a \citet{Calzetti+2000} attenuation curve.}
\vspace{-2mm}
\tablenotetext{$\tiny$ b}{{Derived assuming the median IGM transmission (i.e., the escape fraction values corresponding to 50 \% of the sightlines).}}
\vspace{-2mm}
\tablenotetext{$\tiny$ c}{ $f_{esc}^{abs}$ =  $f_{esc}^{rel} \times 10^{-0.4\,A_{1500}}$, with $A_{1500}$ obtained from the \texttt{Bagpipes} fits.}
\end{deluxetable}

\begin{figure*}[ht]
\includegraphics[width = 0.95\textwidth]{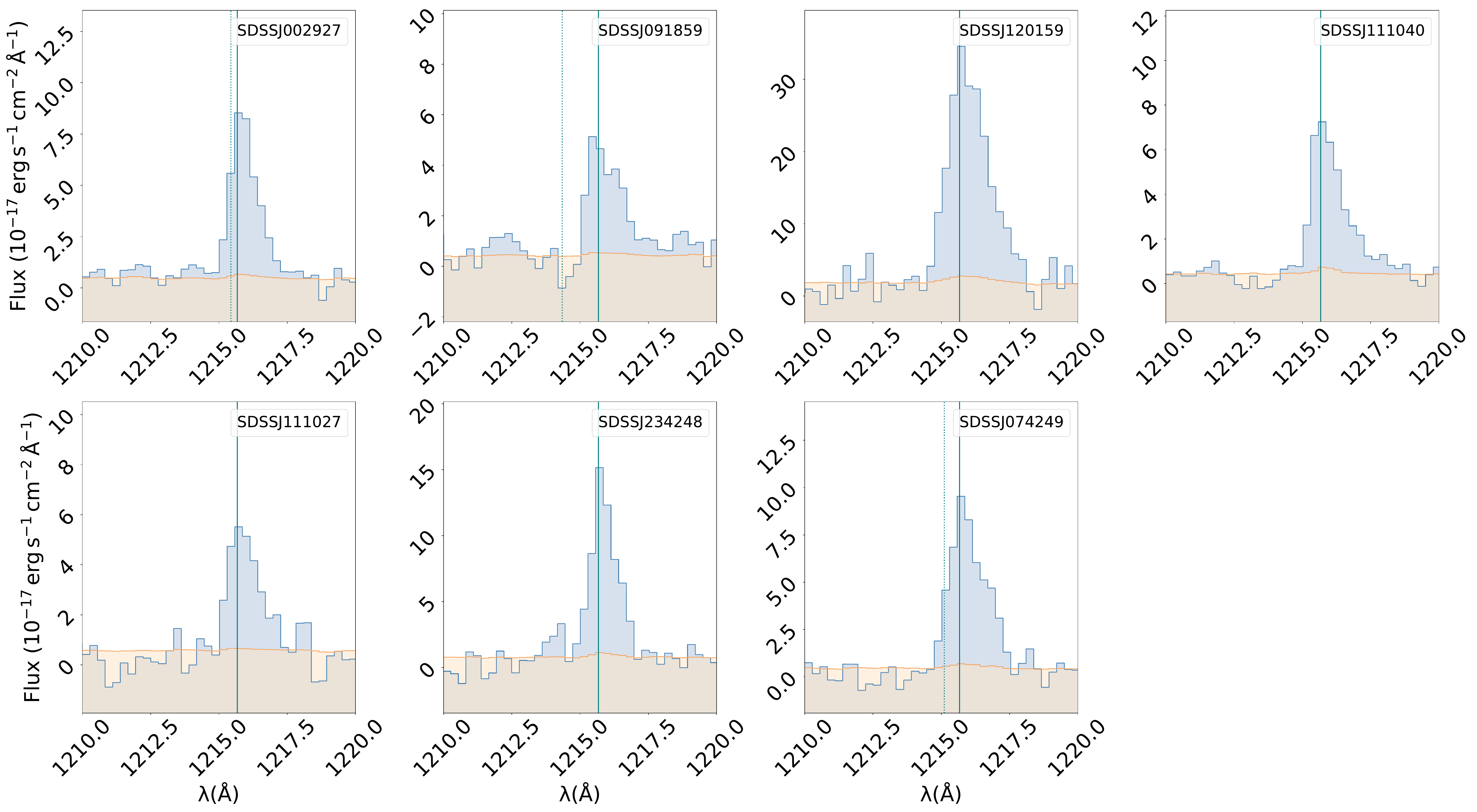}
\caption{{Ly$\alpha$ line profiles in wavelength space. The solid green lines mark the rest-frame wavelength 1215.67 \wa. The green vertical dotted lines mark the wavelength corresponding to $\Delta(v_{Ly\alpha})$ for the three galaxies for which the systemic redshift has been estimated.} The error spectrum is overlayed in orange. {Note that the \lya\ blue peak in SDSSJ234248 is above the noise level.} 
}
\label{fig:lya_lam}
\end{figure*}

\subsection{\texorpdfstring{$Ly\alpha$}{Lya} properties}
\label{subsec:lya_emission}

We measure the \lya\ flux and equivalent width (\ewlya, hereafter) from the BOSS spectra and the HST images. {First, the BOSS spectra are corrected for Milky Way extinction using the \citet{Schlegel+1998} dust map} and the \citet{Cardelli+1989} extinction curve. Subsequently, the continuum of the foreground elliptical galaxy is estimated by fitting the lens subtracted BOSS spectrum using the \textsc{starlight}\footnote{\url{http://www.starlight.ufsc.br/}} full spectrum fitting code \citep{CidFernandes+2005}. We use \citet{BruzualCharlot2003} simple stellar population models, adopting a Chabrier IMF \citep{Charbier2003}, a grid of stellar metallicities Z = 0.0001, 0.0004, 0.004, 0.008, 0.02, 0.05, and ages from 10 Myrs to 3 Gyrs (i.e., the age of the Universe at $z\sim2.3$). After subtracting the continuum from the BOSS spectrum, the Ly$\alpha$ is measured by direct integration of the \lya\ profiles between $\sim$ 1,210 \wa\ and $\sim$ 1,220 \wa.
We use a Monte Carlo approach in order to infer the uncertainties on the flux measurements.

Figure \ref{fig:lya_lam} shows the \lya\ line profiles of our targets. All the galaxies exhibit a prominent \lya\ emission peak, with SDSSJ234248 possibly showing a red and a faint blue peak. We remind  that the observed \lya\ emission is captured within the 2 arcsecond diameter BOSS fiber, which does not, or barely, include the arc structures (see Figures \ref{fig:general1} and \ref{fig:general2}). Therefore, the observed \lya\ emission is not indicative of the total emission within the lensed sources.

The FWHMs of the \lya\ emission range between $250$ and $\sim 400\,\rm km\,s^{-1}$ (corrected for instrumental resolution). It is worth noticing that, for SDSSJ07429 and SDSSJ091859, \citet{MarquesChaves+2020} derived similar FWHMs from low-ionization ISM absorption lines (i.e., Si\thinspace{\sc ii} $\lambda$1260, Si\thinspace{\sc ii} $\lambda$1526 and C\thinspace{\sc ii} $\lambda$1334) within slits positioned on portions of the arc structures. 
Among the seven targets, we have systemic redshift ($z_{sys}$) estimates for  SDSSJ07429, SDSSJ091859, and SDSSJ002927. For SDSSJ07429 and SDSSJ091859, $z_{sys}$ was derived from absorption, nebular and fine-structure emission lines \citep{MarquesChaves+2020}. For SDSSJ002927, we derive $z_{sys}$ from IR spectra observed with the LUCI instrument on the LBT telescope, using strong emission lines ($H\beta$, [O\thinspace{\sc iii}]$\lambda 4959$ and [O\thinspace{\sc iii}]$\lambda5007$) (Citro et al. in prep). The velocity shifts  with respect to systemic range between $\approx 60$  to 300 $\rm km\,s^{-1}$, and are reported in Table~\ref{tab:lya}. 

From the \lya\ fluxes, we compute the observed rest-frame \lya\ equivalent width as:

\begin{equation}
\ W_{Ly\alpha}^{RF} =\frac{\rm Flux_{Ly\alpha}}{\rm F606W} \cdot \frac{1}{(1+z)}~~, 
\end{equation}

where the F606W flux density is used as a proxy of the continuum at 1,216 \wa, and $z$ is the redshift of our galaxies (where possible, we assume $z_{sys}$, otherwhise we use the redshift derived from the \lya\ emission, $z_{Ly\alpha}$ - see Table \ref{tab:properties}). As already mentioned, since the \lya\ is measured within the BOSS spectroscopic aperture, it is likely a lower limit to the total \lya\ flux of each object. Accordingly, in the following Figures and Discussion, we consider these equivalent widths as lower limits. 

\begin{deluxetable}{r|cccc}[h]
\tabletypesize{\footnotesize}
\tablecolumns{5} 
\tablecaption{\label{tab:lya}{Properties of the \lya\ emission line for the seven BELLS GALLERY galaxies analyzed in this work.}}
\tablehead{\colhead{Target} & \colhead{$F_{Ly\alpha, obs}^{~a}$} & \colhead{\ewlyao$^b$} & \colhead{$v_{Ly\alpha, red}^{~c}$} & \colhead{$f_{esc}^{Ly\alpha}$}}
\startdata
SDSSJ002927 & 47.29 $\pm$ 1.12    & 37.52 $\pm$ 0.89 & 58.16 $\pm$ 0.48 & 0.22\\
SDSSJ074249 & 75.75 $\pm$ 2.45 &  68.75 $\pm$ 2.22  & 138.42 $\pm$ 0.30 & 0.39\\
SDSSJ091859 & 104.03 $\pm$ 8.40  &  55.24  $\pm$ 4.46  & 323.8 $\pm$ 2.04 & 0.30\\
SDSSJ111027 & 43.59 $\pm$ 0.40  & 211.91 $\pm$ 1.95 &   - & 1.00\\
SDSSJ111040 & 46.47 $\pm$ 2.16  & 35.99 $\pm$ 1.67 &  - & 0.22\\
SDSSJ120159 & 370.78 $\pm$ 12.35 & 194.78$\pm$ 6.49  & - & 0.94\\
SDSSJ234248 & 78.82 $\pm$ 1.11  &  205.64 $\pm$  2.89 &  - & 1.00\\
\enddata 
\tablenotetext{$\tiny$ a}{Observed-frame fluxes, in units of $\rm 10^{-17}\,erg\,s^{-1}\,cm^{-2}$ (not corrected for the magnification factor).}
\vspace{-2mm}
\tablenotetext{$\tiny$ b}{Rest-frame \lya\ emission equivalent widths, in units of \SI{}{\angstrom}.}
\vspace{-2mm}
\tablenotetext{$\tiny$ c}{Velocity shift of the red peak of the \lya\ emission with respect to the systemic velocity, in $\rm km\,s^{-1}$. We show this value only for galaxies with an estimate of the systemic redshift.}
\end{deluxetable}

\section{Results and Discussion}
\label{sec:results}

\subsection{Escape fraction vs. galaxy properties} 
\label{sec:fesc_uvmag}

In the following, we explore the relationships between the LyC escape fraction of our targets and their $\beta$ slope, $M_{UV}$ (from the F606W images), \ewlyao, and \lya\ escape fraction. {In subsection \ref{subsec:lyc_emission}, we derived that our targets are not LyC emitters, with LyC escape fractions \fa $\lesssim 13$ \% when the median IGM transmission is considered. In this regard, we stress that this is not the most conservative estimate of \fa. In fact, if 84 \% of the sightlines are considered, our \fa\ upper limits become more stringent, decreasing to $\lesssim 6$ \% (see Figure \ref{fig:fesc_igm})}.
Moreover, our findings of subsection \ref{sec:dust_ext} indicate that our targets have $\beta$ ranging between $-2.4$ and $-2.0$, and are consequently relatively low in dust content.
Our median value of $\beta$ aligns with findings from other studies on faint galaxies \citep[e.g.,][]{Venemans+2005, Nilsson+2007, Ouchi+2008, Hayes+2010, Ono+2010a, Nakajima+2012, MattheeSobral2016}.
The low dust content implied by the blue $\beta$ slopes are confirmed by the \texttt{Baagpipes} fits to the photometry, which indicate $E(B-V)_{\rm fit}$ ranging {between 0.04 to 0.1 (see Table \ref{tab:betaslope}).}

\begin{figure}
\includegraphics[width=\columnwidth]{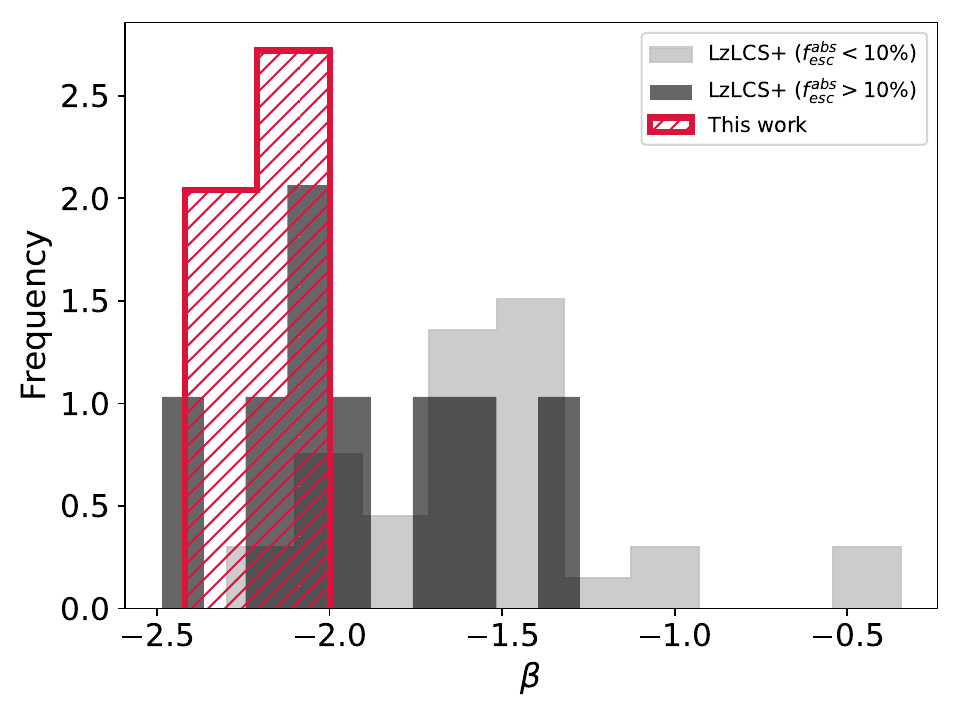}
\caption{$\beta$ frequency distribution for our targets and LzLCS+ targets (i.e., LzLCS galaxies from \citealp{Flury+2022a,Flury+2022b} combined with HST/COS data from \citealp{Izotov+2016a,Izotov+2016b, Izotov+2018a, Izotov+2018b, Wang+2019}). The silver histogram represents the $\beta$ distribution for local galaxies with \fa\ $<10$ \% at 1$\sigma$ significance. The black histogram shows the distribution for local galaxies with \fa\ $>10$ \% at 1$\sigma$ significance. The red histogram illustrates the $\beta$ distribution for our sample (\fa\ $ \lesssim 13$ \% when the median IGM transmission is considered). }
\label{fig:chisholm}
\end{figure}

Figure \ref{fig:chisholm} shows the comparison between the distribution of the $\beta$ slopes of our targets and that of the low-redshift LzLCS+ sample (i.e., LzLCS galaxies from \citealp{Flury+2022a,Flury+2022b} combined with HST/COS data from \citealp{Izotov+2016a,Izotov+2016b, Izotov+2018a, Izotov+2018b, Wang+2019} and \citealp{Izotov+2021}). It is evident that our galaxies show $\beta$ values that overlap with the LzLCS+ distribution. However, at $\beta\lesssim-2$, the majority of the local galaxies are characterized by \fa\ $> 10$ \% at 1$\sigma$ significance. For similar $\beta$, our galaxies show {instead \fa\ $\lesssim 13$ \% when the median IGM transmission is considered}. Our results are similar to what observed by \citet{Jung+2024} in galaxies with $-18\lesssim M_{UV} \lesssim-21$ at $z\sim1.3-3$.


\begin{figure*}[ht!]
    \includegraphics[width = \textwidth,  trim={0.5em 12em 0.2em 10em}, clip]{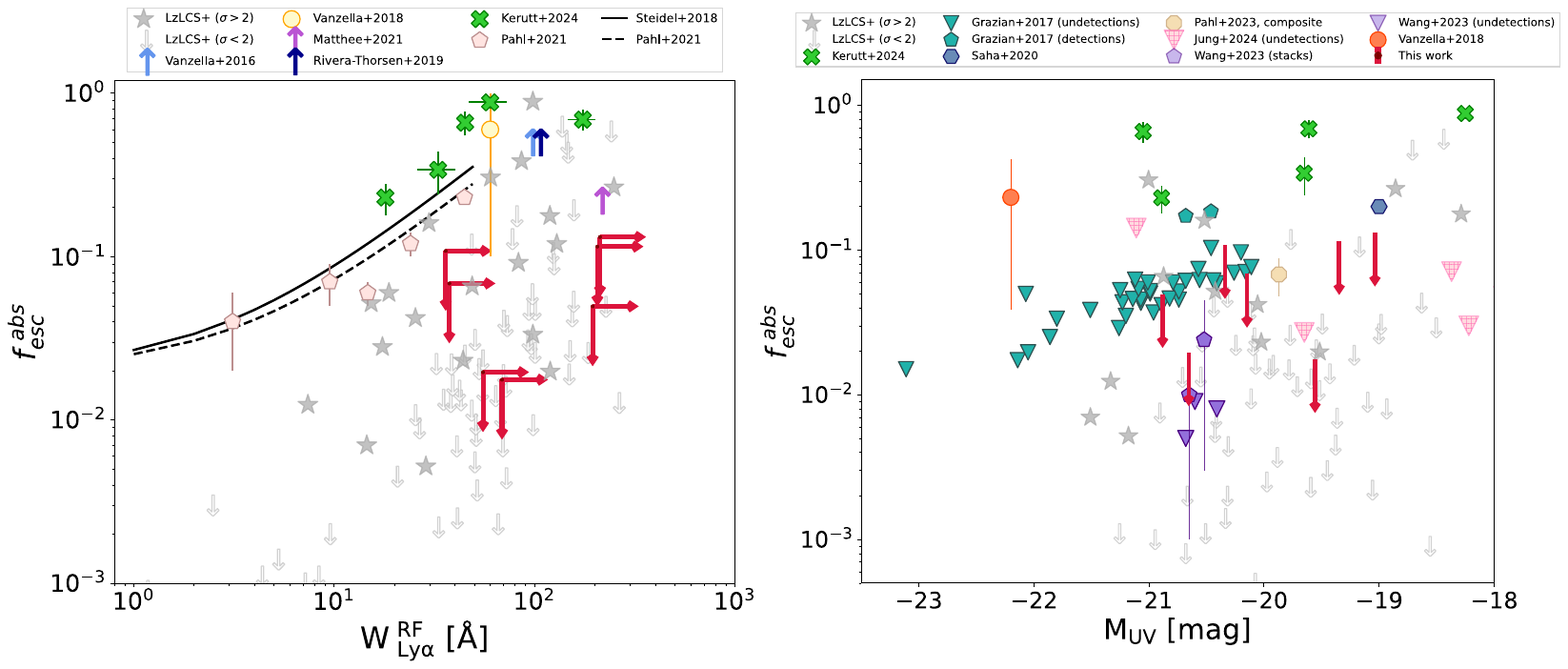}
\caption{{Left panel}: \fa vs. \ewlya\ plane. Black and grey symbols are measurements in the local Universe at different detection levels (stars - $>2\sigma$, downward pointing arrows $<2\sigma$) from the LzLCS+ sample. High redshift LyC leakers are marked with colored symbols. Specifically, we show \textit{Ion2} \citep{Vanzella+2016}, \textit{Ion3} \citep{Vanzella+2018}, the \textit{Sunburst Arc} \citep{Rivera-Thorsen+2019}, GS 30668/XLS-26 \citep{Naidu+2017, Matthee+2021a}. We also show results from \citet{Pahl+2021} and \citet{Kerutt+2024}.
Red downward and rightward pointing arrows are the \fa\ upper limits and the lower limits on the \ewlyao\ for the seven BELLS GALLERY galaxies studied in this work.  {Right panel}: absolute escape fractions as a function of the UV magnitude at 1,500 \wa. Black and grey symbols are color coded as in the left panel. Red downward pointing arrows are the  \fa\ upper limits for the seven BELLS GALLERY galaxies studied in this work. Downward pointing triangles of different colors are undetections from \citet{Grazian+2016} (1$\sigma$), \citet{Wang+2023} (2$\sigma$), and \citet{Jung+2024} (1$\sigma$). Other symbols are detections from \citet{Grazian+2016}, \citet{Wang+2023}, \citet{Vanzella+2018}, \citet{Saha+2020}, \citet{Pahl+2023}, and \citet{Kerutt+2024}, respectively (when relative escape fractions are provided, we transform them into absolute escape fractions by multiplying assuming a \citet{Calzetti+2000} attenuation curve and  $\rm E(B-V)=0.1$).}
\label{fig:fesc_composite}
\end{figure*}

Figure \ref{fig:fesc_composite} shows \fa\ as a function of \ewlya\ (left panel) and $M_{UV}$ (right panel) for our galaxies and compiled literature data at various redshifts. {The data shown in this Figure are all obtained adopting either a \citet{Calzetti+2000} or a \citet{Reddy+2016} extinction curve, which are similar. However, as explained in subsection \ref{subsec:lyc_emission}, we also perform our \fa\ calculations adopting the SMC extinction curve.  For context, the Calzetti curve yields higher $E(B-V)$ values and consequently lower \fa\ compared to the SMC curve. Therefore, all data in Figure \ref{fig:fesc_composite} would shift towards \fa\ values higher by a factor $\sim$ 1.4 if the SMC curve were used.}
Studies in the local Universe (grey symbols in the Figures) extend the measurements to faint UV magnitudes and escape fraction measurements/limits down to sub-percent values. Conversely, high redshift studies are typically limited to larger values of the escape fraction \citep[e.g.,][]{Grazian+2016, Vanzella+2016, Jung+2024, Kerutt+2024}. Our sample provides a stringent upper limit on the escape fraction of LyC radiation down to $M_{UV}=-19$ at $z\sim2.3$.

In Figure \ref{fig:fesc_composite} (left panel) { we observe that the \ewlyao\ of the BELLS GALLERY galaxies range from  {$\sim$ 40 to $\sim$ 200 \wa}, which corresponds to the higher envelope of \ewlyao\ observed in the LzLCS+}. In this range, most of the LzLCS+ galaxies are not LyC leakers, which would be consistent with the observed non detections. However, as mentioned earlier, the \ewlya\ of our $z\sim 2$ galaxies should be considered as lower limits. Given what is known about the extent of \lya\ emission in both local and high redshift galaxies \citep[e.g.,][]{Hayes+2013, Leclercq+2017, Erb+2018, Claeyssens+2022, Erb+2023}, it is entirely plausible that the \lya\ emission within the BOSS fiber is largely underestimating the real emission. 
To quantify this effect, we estimate the total \lya\ flux using relations based on low redshift galaxies. Specifically, we estimate the total \lya\ flux by multiplying the \ewlyap\ derived from the \texttt{Bagpipes} fits by the UV continuum in the F606W band. We then assume that this flux is distributed in 9 $\times$ area of the UV continuum (following \citealp{Hayes+2013}). {From this calculation, we derive that the BOSS fiber contains only up to $\sim40$ \% of the actual emission, bringing the actual \ewlya\ range to $\sim$ 100 - 500} \wa. For galaxies with \ewlya\ $>$ 100 \wa, {we would expect more than 50 \% of our galaxies (i.e., 3-4) to be LyC emitters, based on the results from \citet{Flury+2022b}  at $z=0.3$. We acknowledge that comparing our non-detections with results from high-redshift samples would be more appropriate, as high-redshift sources are significantly more affected by IGM absorption than local ones. However, such a comparison is challenging due to the inconsistent LyC detection rates reported in the literature, which vary from 0 \% \citep[e.g.,][]{BianFan2020} to 20 \% \citep[e.g.,][]{Fletcher+2019} at $z\sim 3$. Moreover, an understanding of how the LyC detection rate evolves with \ewlya\ at high redshift is still lacking, making it difficult to draw a meaningful comparison with our non-detections at high \ewlya. We refer to Section \ref{sec:igm_abs} for a more detailed discussion on how the foreground IGM is affecting the LyC emission of our targets. Here we highlight that the key point of our analysis is that our non-detections are coupled with surprisingly high \ewlya\ compared to other literature studies.} 

{Figure \ref{fig:fesc_composite} (left panel) also shows that our \fa\ upper limits lie below the \fa\ vs. \ewlya\ relation  defined by \citet{Steidel+2018} and \citet{Pahl+2021}. If this relationship reflects the neutral hydrogen covering fraction influencing both
 \lya\ emission and LyC leakage, then our galaxies must be characterized by a gas covering fraction $<1$, which allows \lya\ photons to escape. However, optically thick clumps
(with $N_{HI} > 10^{17}$) must be present to absorb most of the ionizing photons, similarly to what described in \citet{Gazagnes+2020}.}


{On the right panel of Figure \ref{fig:fesc_composite}, we observe that our sample of galaxies has upper limits on \fa\ that are generally compatible with those found in the local and high redshift Universe \citep[e.g.,][]{Grazian+2016, Jung+2024}. The fact that the discrepancies in \fa\ are more pronounced when considering \ewlya\ rather than $M_{UV}$ highlights the complexity of the relation between LyC and \lya\ emission, and especially the role of the ISM gas covering fraction, which affects \ewlya\ but not $M_{UV}$.} {It is interesting to note that our LyC upper limits are considerably lower than the high-$z$ detections of \citet{Kerutt+2024}, although having similar $M_{UV}$. 

{In Figure \ref{fig:pahl}, we plot \fa\ against the velocity separation between the red Ly$\alpha$ peak and systemic velocities for our galaxies, local green peas \citep{Kakiichi+2021}, and stacks at $z\sim3$ \citep{Pahl+2024,Kerutt+2024}. The $z\sim 2.3$ lensed Ly$\alpha$ emitters exhibit lower \fa\ than those measured in the local Universe and in $z\sim3$ stacked spectra for similar velocity shifts, indicating a suppression of \fa\ at high redshift. This contrasts with the findings by \citet{Pahl+2024}, who observed higher \fa\ for a given Ly$\alpha$ velocity shift compared to low-$z$ studies. We do not discuss this discrepancy further as we only have three galaxies with individual velocity shift measurements. However, this finding aligns with a scenario where high-redshift galaxies have larger amounts of neutral hydrogen. Notably, targets SDSSJ002927 and SDSSJ074249 have Ly$\alpha$ velocity shifts $\lesssim$ 150 $\rm km,s^{-1}$. In the common understanding of Ly$\alpha$ scattering, velocity shifts smaller than $\approx 150$\,km\,s$^{-1}$ indicate low column density of neutral gas \citep{Verhamme+2015},  which is at odds with the non-detection in LyC. These two targets might have a gas covering fraction $<1$, i.e. a low average neutral hydrogen density, but high column density clumps that absorb LyC photons along certain paths. This scenario is compatible with what recently inferred by \citet{Li+2024} using radiative transfer modeling on local LyC emitters. Even though our trends are opposite to those seen by \citet{Pahl+2024}, we reach a similar conclusion that caution should be used when employing relations between the Ly$\alpha$ profile and \fa\ calibrated at low-$z$ as indicators of \fa\ at high redshifts.}

\subsection{Absorption of LyC by the intergalactic medium}
\label{sec:igm_abs}
{In this subsection, we analyze in more detail how the IGM is affecting the LyC radiation emitted by our targets and discuss its properties. To do this, we start with calculating the angle formed by the undisturbed light rays coming from the source. This angle subtends the distance between the lensed images around the lens, essentially representing the spread of light rays from the source. For our galaxies, we find that this angle is, on average, $1''$. We then compare this angle to the typical angular size of neutral hydrogen absorbers with $N_{HI} \approx 10^{17}$ at $z\sim2.3$. 
Using Equation 12 from \citet{Schaye2001}, we first derive a typical physical size of 10 kpc. Since the absorbers could be located anywhere between the source and the lens, we determine the angular size of the absorbers for various redshifts between $z_{lens}$ and $z_{Ly\alpha}$ (see Table \ref{tab:properties}). We obtain values $\lesssim\,1''$. This suggests that the LyC emission from each one of our seven galaxies might be absorbed by a single, uniform patch of IGM rather than by multiple small absorbers. However, it is crucial to remember that the sightlines to our seven galaxies are independent. Specifically, from the IGM transmission distributions obtained in Section \ref{subsec:lyc_emission}, we find that approximately 15 \% to 30 \% of the 500 simulated sightlines have IGM transmission below 20 \%. This implies that the likelihood of all seven galaxies having optically thick absorbers along their sightlines is $\sim$ 0.02 \%. Additionally, the fact that the sightlines are separated by 10s of degrees in the sky minimizes the effect of correlated \lya\ forest absorption recently identified by \citet{Scarlata+2025}. These considerations suggest that for at least some of our galaxies, the observed LyC non-detections are intrinsically related to ISM/CGM properties rather than solely to IGM absorption. However, {increasing the statistics} and obtaining additional data, such as optical spectra and high-resolution \lya\ profiles, would be necessary to further investigate the stellar and ISM properties of our sources and draw more definitive conclusions.}

\begin{figure}[h]
    \includegraphics[width = \columnwidth, trim={0.2em 0em 0.em 0em}, clip]{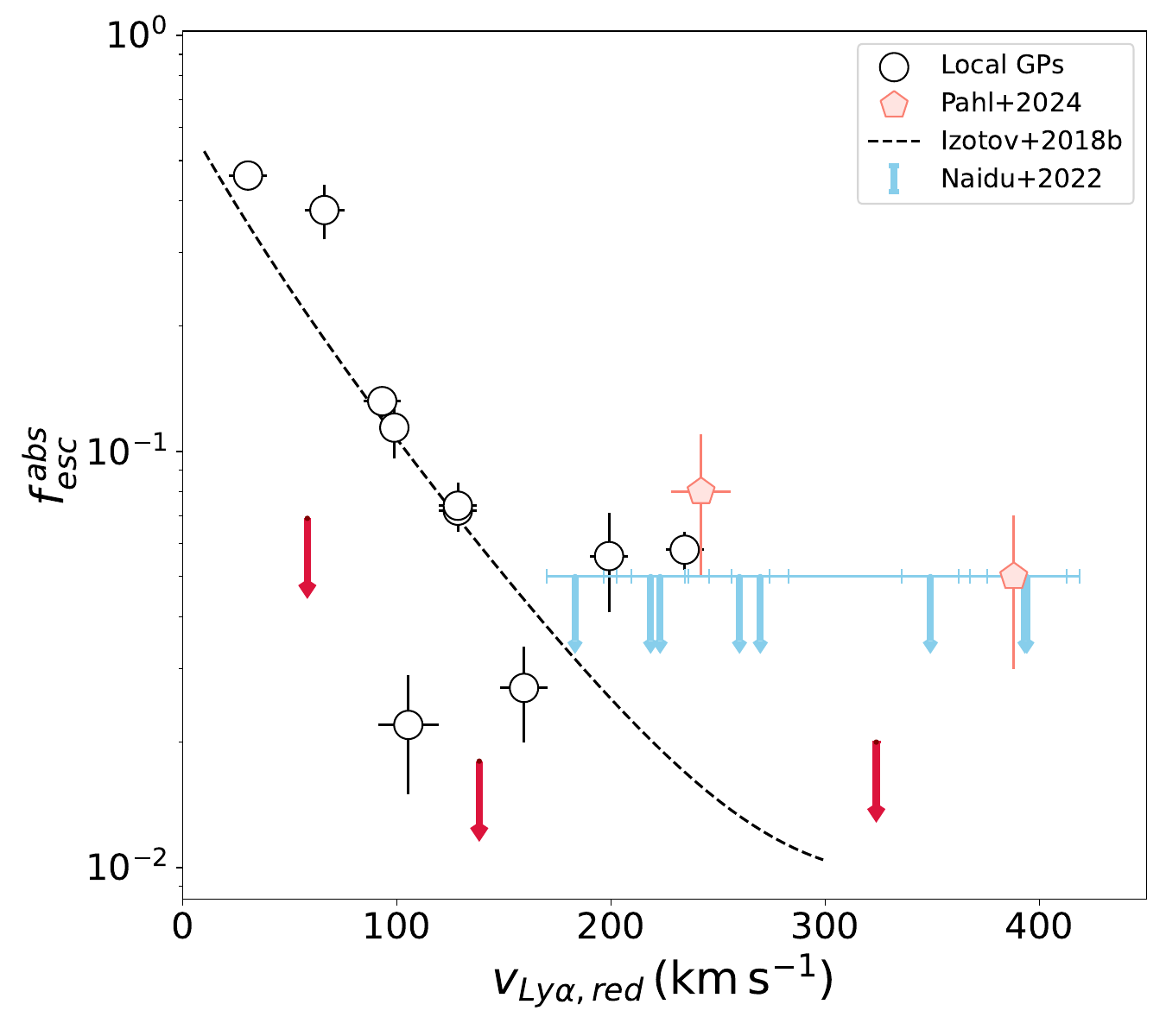}
\caption{{Absolute escape fraction versus \lya\ red peak velocity ($v_{Ly\alpha,red}$) for galaxies at different redshifts. Red arrows are the results obtained in this work for the three galaxies for which we have systemic velocity measurements (see Table \ref{tab:lya}); white circles are local GPs measurement summarized in \citet{Kakiichi+2021}; Pink pentagons are high redshift measurements from \citet{Pahl+2024}. Blue downward arrows are the upper limits by \citet{Naidu+2022}.
Note that where only $v_{sep}$ (the separation between the red and blue peaks of the \lya\ emission) was available, $v_{sep}$ was converted to $v_{Ly\alpha, red}$ using Equation 8 from \citet{Pahl+2024}.}}
\label{fig:pahl}
\end{figure}

\subsection{An analytical approach to model the LyC and \texorpdfstring{$Ly\alpha$}{Lya} radiative transfer}

\label{sec:factors}

\begin{figure*}
\centering
\includegraphics[width=0.8\linewidth, trim={1em 5em 1em 5em}, clip]{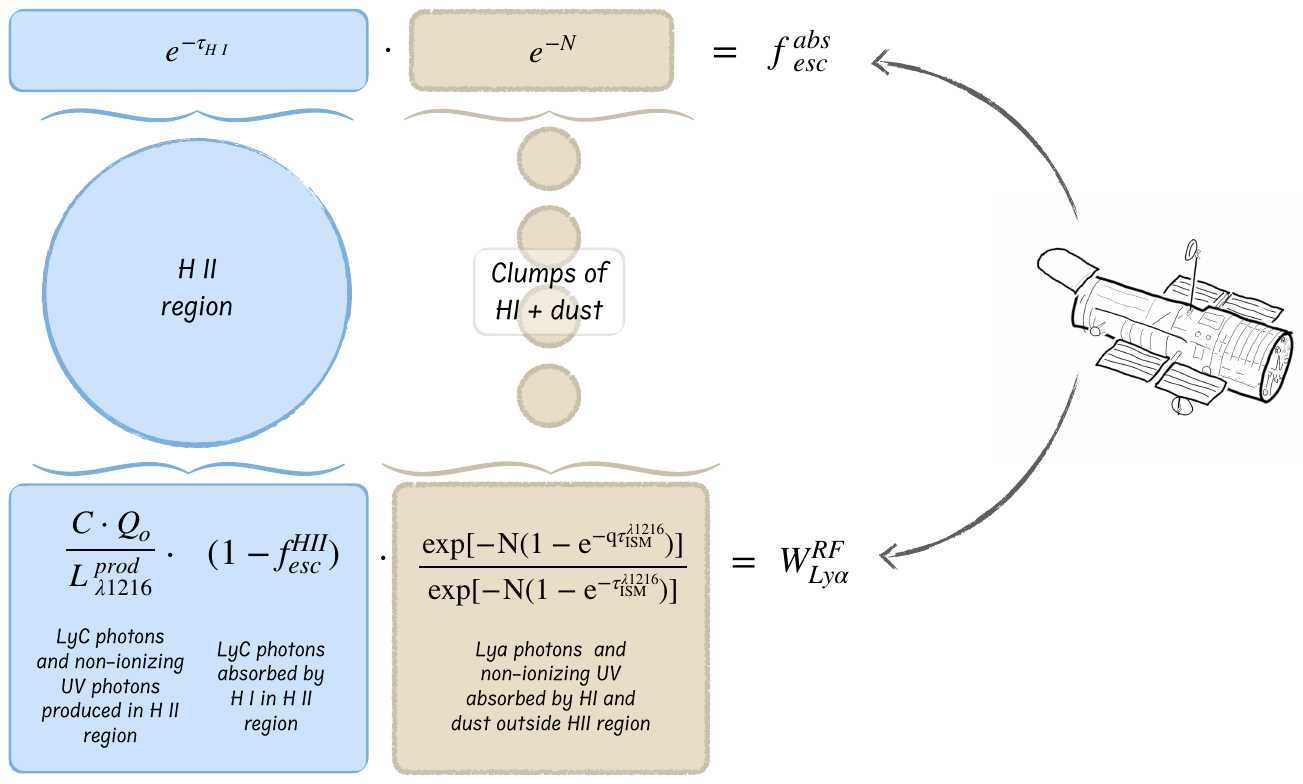}
\caption{Schematic view of our models and \ewlyao\ parameterization in the case of a dust screen positioned ahead of an \ion{H}{2} region, the source of \lya\ photons. It is important to highlight that the observed escape fraction \fesco\  consistently remains lower than the escape fraction at the outer edge of the \ion{H}{2} region ($f_{~esc}^{~H\,II}$), as LyC photons can be absorbed by the neutral hydrogen and dust within the clumpy dust screen.}
\label{fig:models}
\end{figure*}

In this subsection, we develop a simple toy-model to describe the effects of radiative transfer of \lya\ and non-ionizing UV photons on the resulting rest-frame \ewlyao. We assume that young stars produce a spherical \hii\ region, free of dust, where \lya\ is produced after recombination. The star light and the \lya\ photons then travel through a clumpy screen of dust and \hi.
In this model, schematically shown in Figure \ref{fig:models}, the observed \ewlyao\ depends on the value of the equivalent width of the \lya\ at the outer edge of the \hii\ region ($W_{Ly\alpha}^{H\,II}$) and on the radiative transport of \lya\ and continuum photons through the clumpy medium. $W_{Ly\alpha}^{H\,II}$ depends on 
the intrinsic \ewlyap\ (i.e., resulting from age, metallicity and IMF-dependent stellar population -- assuming all ionizing photons are absorbed), and corrected for the fraction of ionizing photons that escape the \hii\ region. Accordingly:


\begin{equation}
W_{Ly\alpha}^{H\,II} = W_{Ly\alpha}^{prod} \cdot (1-f_{esc}^{H\,II}) = \frac{C\cdot Q_{0}}{L_{\lambda1216}^{prod}} (1-f_{esc}^{H\,II}),
\label{eq:ewlya_prod}
\end{equation}

where $Q_{0}$ is the number of ionizing photons, $C=1.04\times10^{-11}$\,erg is the \lya\ emission coefficient for Case B recombination, $n_{e} = 100$\,cm$^{-3}$  (see \citealp{Schaerer2003}),  $L_{\lambda1216}^{prod}$ is the continuum luminosity  at 1,216 \wa, and $(1-f_{esc}^{H\,II})$ is the fraction of LyC photons absorbed in the \ion{H}{2} region and resulting in \lya. Note that $f_{esc}^{H\,II}$ is not the global LyC escape fraction  measured at Earth, as ionizing photons that are not used to produce nebular lines still need to traverse the galaxy's ISM/CGM, where they can be absorbed by dust and neutral hydrogen.

According to the chosen geometry, the attenuation due to the clumps can be written as $exp[{-N(1-e^{-(\tau_{ISM,c}^{\lambda912}+\tau_{HI,c})})]}$, where $N$ is the average number of clumps along the line of sight  \citep[e.g.,][]{Scarlata+2009}. Given that $(\tau_{ISM}^{\lambda912}+\tau_{HI})>>1$, (with  $\tau_{ISM}^{\lambda912}<<\tau_{HI}$), this expression can be approximated as  $e^{-N}$, i.e., the covering fraction of the clumps. \lya\ photons are not destroyed when absorbed by \hi, but rather re-emitted along different directions and with a frequency shift that depends on the gas motions, resulting in an excess probability (with respect to non-resonant photons) of being absorbed by dust. This effect can be modeled using the ``scattering $q-$parameter" introduced by  \citet{Scarlata+2009} and \citet{Finkelstein+2009}. The resulting attenuation of \lya\ photons is $exp[-N(1-e^{-q\tau_{ISM}^{\lambda1216}})]$.

\begin{figure*}[ht!]
\includegraphics[width = \textwidth, trim={1em 12em 1em 5em}, clip]{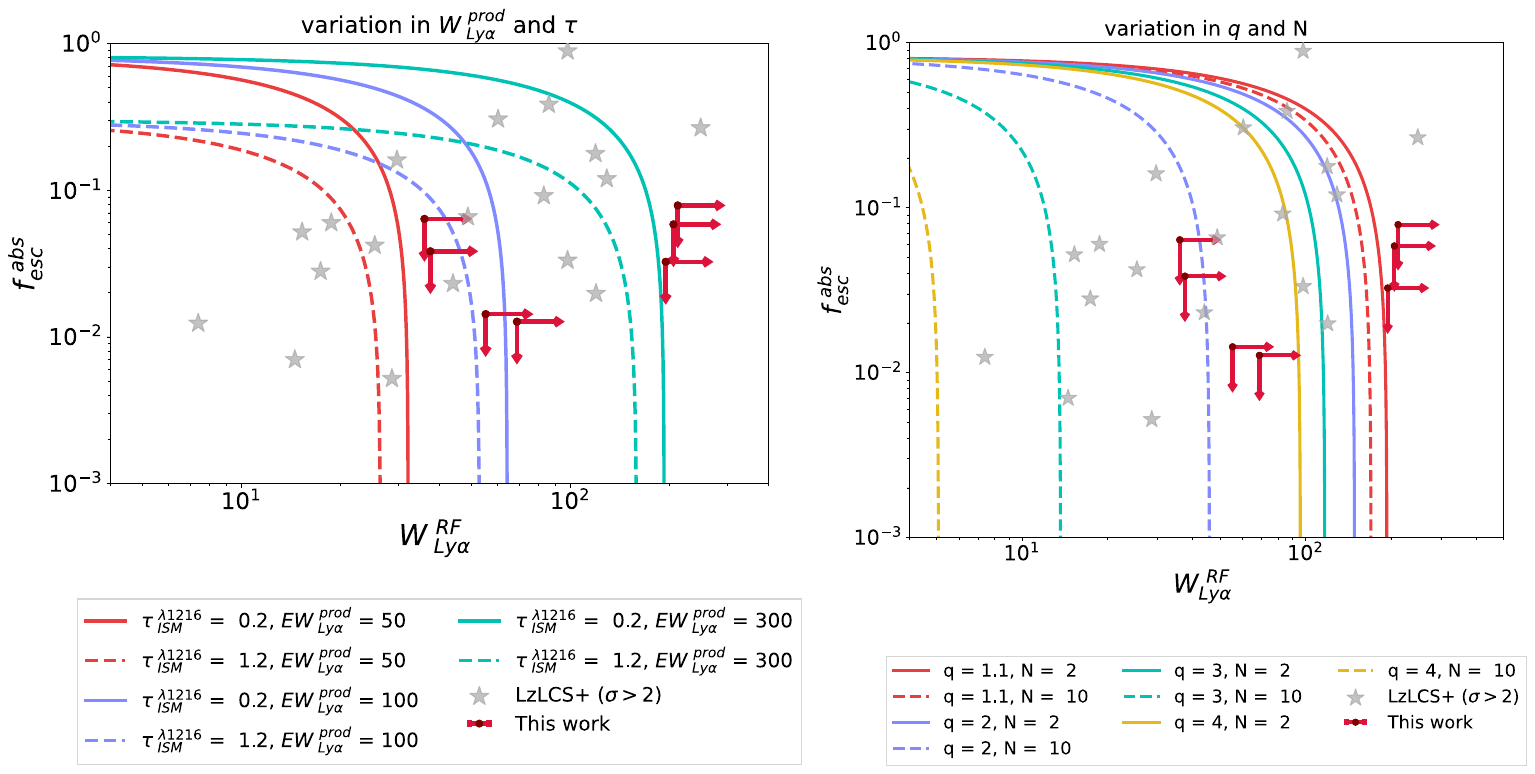}
\caption{Toy model curves overimposed to the LzLCS+ ($\sigma>2$) (grey stars) and our seven targets (red arrows). {Left panel:} models vary in \tauism\ and \ewlyap\ according to the colors and line styles described in the legend. Values of $q=2$ and $N = 3$ are assumed. {Right panel:} models vary in $q$ and $N$ according to the colors and line styles described in the legend. Values of \ewlyap\ = 200 \wa\ and \tauism = 0.2 are assumed.}
\label{fig:models_ew_q_var}
\end{figure*}

To summarize, \ewlyao\ can be written as:

\begin{equation}
W_{Ly\alpha}^{RF} =  W_{Ly\alpha}^{prod} \cdot (1-f_{esc}^{H\,II}) \cdot \frac{exp[-N(1-e^{-q\tau_{ISM}^{\lambda1216}})]}{exp[-N(1-e^{-\tau_{ISM}^{\lambda1216}})]}, 
\label{eq:final}
\end{equation}

where $\rm \tau_{ISM}^{\lambda1216}$ is the dust optical depth at 1,216 \wa.

Figure \ref{fig:models_ew_q_var}  shows the analytical models overimposed to our data and the LzLCS+ results with $> 2\sigma$ significance.  We observe that an increase in \tauism\ leads to a decrease of both \fa\ and \ewlyao. Additionally, decreasing \ewlyap, increasing the number of clumps $N$, or increasing the scattering parameter $q$ leads to a decrease in \ewlyao. The LzLCS+ galaxies with higher \fa\ can be modeled with a low \tauism, a high \ewlyap, a low $q$, and a low $N$. It is interesting to note that the parameters required to reproduce the LzLCS+ galaxy with \fa $\sim90$ \% are particularly extreme, involving either an \ewlyap $>1000$ \wa, a \tauism $<0.05$, or nearly zero covering fractions. Instead, galaxies with lower \fa\ and \ewlyao\ are generally characterized by either a lower \ewlyap, a higher $q$, or a higher $N$. Unfortunately, our models are degenerate, with different combinations of the parameters producing similar values of \ewlyao\ and \fa. Currently, we lack the necessary information to break this degeneracy. Independent measurements, such as the H$\alpha$ and H$\beta$ lines, could help determine the covering fraction of the clumps $e^{-N}$ \citep{Scarlata+2009} which, combined with the derived \tauism, \fa, and \ewlyap, would allow us to estimate $q$ and understand the degree of scattering that \lya\ photons experience on their way out of the galaxy.



\subsection{\texorpdfstring{$Ly\alpha$}{Lya} escape fraction}
\label{sec:explanation}
Using the results from the SED fitting discussed in subsection~\ref{sec:fesc_uvmag}, we can compute the escape fraction of \lya\ radiation, that we define as \ewlyao/\ewlyap {(the obtained values are listed in Table \ref{tab:lya})}. \citet{Flury+2022b} show that there is a strong correlation in the local universe between the \lya\ and LyC escape fractions. 

The \lya\ and LyC escape fractions are compared in Figure~\ref{fig:flya}.
The $z\sim 2.3$ galaxies studied here have higher \lya\ escape fractions than the low-$z$ galaxies' distribution, result that is even stronger when considering that the observed \lya\ fluxes are lower limits (see subsection \ref{sec:fesc_uvmag}). Given the strong correlation between the \lya\ and LyC escape fractions in the local universe, the \lya\ escape fraction is expected to be a reliable indicator of the the LyC escape fraction. How, then, can we reconcile this with the LyC non-detections and the strong upper limits suggested by our measurements?

We suggest that this discrepancy may be due to an increasing neutral gas fraction for higher redshift/lower mass galaxies \citep[e.g.,][]{Chen+2021, Papastergis+2012, Maddox+2015}, accompanied by a progressively lower dust content \citep[e.g.,][]{Cucciati+2012, Hayes+2011, Eales+2024}. This would follow naturally given that the dust-to-gas ratio is observed to decrease with both metallicity \citep[e.g.,][]{RemyRuyer+2014} and stellar mass \citep[e.g.,][]{Santini+2014}, a 
trend that seems to hold well into the reionization epoch \citep[e.g.,][]{Heintz+2024}.
At high redshifts, larger neutral hydrogen reservoirs and  reduced dust attenuation would result in a stronger absorption at LyC, while \lya\ radiation would be less affected, despite enhanced scattering in neutral gas. This scenario could be tested via measurements of the \lya\ line profiles, as radiative transfer models suggest that large \hi\ column densities produce broad \lya\ profiles, with large shifts with respect to systemic. With only velocity shifts for three objects, we cannot {fully} explore this hypothesis at this point. {However, we observe that the three galaxies for which we have Ly$\alpha$ velocity shifts exhibit \fa\ upper limits that are lower than those of most local galaxies with similar $v_{Ly\alpha, red}$ (see Figure \ref{fig:pahl}). Despite this, they show a small shift of Ly$\alpha$ compared to the systemic redshift ($v_{Ly\alpha,red}<150,\rm km\,s^{-1}$), compatible with optically thin gas. We believe that these galaxies might be characterized by a gas covering fraction $<1$ with optically thick clumps ($N_{HI}>10^{17}\,\rm cm^{-2}$). In this case, the optically thick clumps suppress the LyC emission, but Ly$\alpha$ photons can escape through clear channels.} This scenario is also consistent with the ``Low escape" phase hypothized by \citet{Naidu+2022}. In their model, channels created by stellar feedback may be refilling with \hi, making it difficult for LyC photons to escape while still allowing \lya\ photons to do so, if the dust content is low.  However, this evolutionary phase would be relatively short-lived (approximately $10-100$ Myrs), making it unlikely that all of our galaxies are currently experiencing it. 

Ultimately, we cannot rule out that our findings are influenced by how BOSS captures the \lya\ emission in the extended halo, whereas LyC emission is measured at the sites of continuum production. In fact, the observed \lya\ emission might have undergone a different radiative transfer compared to the LyC emission. Spatially resolved maps of the \lya\ emission are needed, alongside the existing LyC leakage maps, in order to confirm the true nature of these galaxies and test our hypotheses. 

\begin{figure}[H]
\includegraphics[width=0.95\columnwidth]{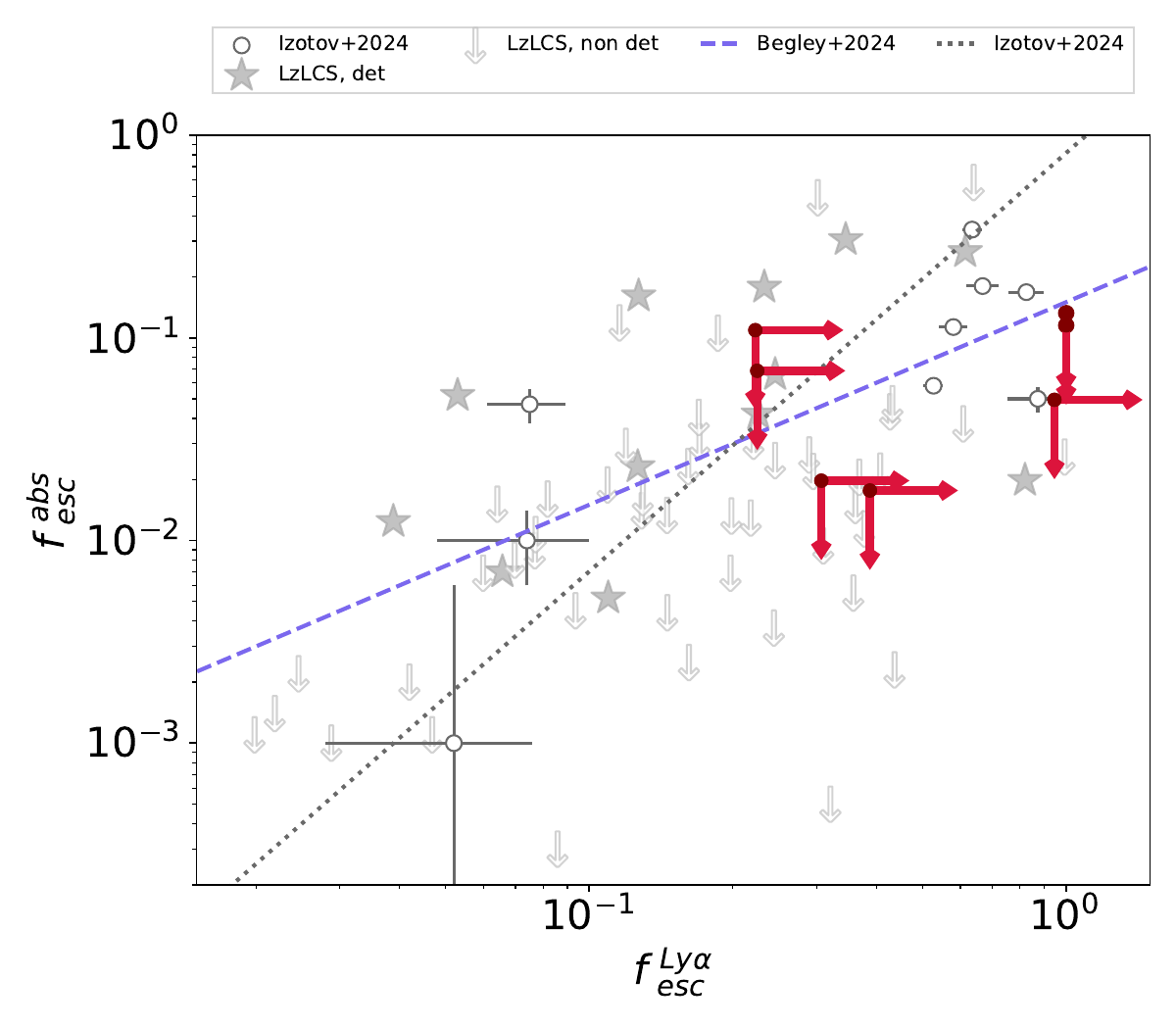}
\caption{\fesco\ vs. \flya. Symbols are color coded as in Figure \ref{fig:fesc_composite}. {Note that, for galaxies IDs SDSSJ111027 and SDSSJ234248 we do not plot the lower limits in \flya\ since they already have \flya $=1$.} The purple dashed and orange dotted lines are the \fa\ vs. \flya\ relations by \citet{Begley+2024} (VANDELS galaxies) and \citet{Izotov+2024} ($z<0.5$), respectively.}
\label{fig:flya}
\end{figure}

\section{Conclusions}
\label{sec:conclusions}
In this paper, we investigate the relation between the LyC leakage and \lya\ emission in a sample of seven $z\sim2$ gravitationally lensed \lya\ emitters drawn from the BELLS GALLERY Survey \citep{Shu+2016a, Shu+2016b}. Our galaxies {are located at the faint end of the UV luminosity function at $z\sim2-2.5$} (with $-21\lesssim M_{UV}\lesssim-19$), and are dust poor (with $\beta$ ranging between $-2.4$ and $-2.0$). From the UV broad band F275W and F225W HST filters, we provide direct upper limit measurements of the emission below 912\AA, from which we estimate upper limits on \fa. We also use the BOSS spectra to measure lower limits on the \lya\ fluxes and \lya\ equivalent widths. We create a simple analytical model to describe the connection between \lya\ emission, LyC leakage, \ion{H}{1} column density, dust, and scattering effects. 
We find that the seven BELLS GALLERY galaxies exhibit similar $\beta$ slopes as nearby galaxies. However, while local sources show \fa $\gtrsim10$ \% at $1\sigma$ significance, 
{our high redshift targets have \fa $\lesssim13$ \%, based on the median IGM transmission obtained over 500 simulated realizations through the IGM for each of our targets.}

{We compare the derived \fa\ with  \ewlya\ and the shape of the \lya\ profile. We conclude that our galaxies are likely characterized by a gas covering fraction $<1$ with optically thick clumps ($N_{HI} > \rm 10^{17}\,cm^{-2}$). This configuration allows \lya\ photons to escape, while absorbing most of the LyC radiation. From a detailed investigation of the IGM, we conclude that, if the  LyC emission from our targets is attenuated by foreground IGM, it is probabily absorbed by a single patch of IGM instead of many small absorbers. However, we highlight that the probability of all seven galaxies to have foreground IGM with transmission $<20$ \% is $<0.02$ \%. Therefore, we believe that the LyC non-detections are related to intrinsic properties of the ISM, rather than solely to the IGM absorption.}
{Our findings seem to reflect the observed increase in neutral hydrogen and decrease in dust content in high-redshift galaxies, and also the presence of clumpy, optically thick gas at high redshift.} The combination of these factors reduces the escape fraction of LyC radiation, while still enabling \lya\ emission to escape, despite the large number of scatterings. 
Our result implies that the LyC leaking mechanisms could work differently at high redshifts, and therefore that using LyC low-redshift indirect estimators to interpret the high redshift Universe might be incorrect. 
However, additional independent data, such as spatially resolved \lya\ imaging within the arc structures, high resolution \lya\ profiles and optical stellar spectra are necessary to confirm our hypothesis. {Furthermore, Euclid \citep{Euclid2024} will enable us to increase the statistics of gravitationally lensed galaxies to validate our findings.}






\begin{acknowledgments}
{The authors thank the anonymous referee for the insightful suggestions and feedback that improved the manuscript. A.C thanks: Dr. Sophia Flury for providing the LzLCS catalog; Dr. Alexandra Le Reste and Dr. John Chisholm for stimulating and useful discussion;  Dr. Sean Bruton for help with running the SED fits.} A.C and C.S. thank Dr. Adrea Grazian for providing the LUCI spectrum for SDSSJ002927. KBM acknowledges partial funding from the U.S. National Science Foundation Award IIS2006894 and NASA Award 80NSSC20M0057. M.J.H. acknowledges support from the Swedish Research Council, Vetenskapsr$\rm \mathring{a}$det, and is fellow of the Knut \& Alice Wallenberg Foundation. 
Based on observations with the NASA/ESA Hubble Space Telescope obtained at the Space Telescope Science Institute, which is operated by the Association of Universities for Research in Astronomy, Incorporated, under NASA contract NAS5-26555. Support for Program number \href{https://archive.stsci.edu/proposal_search.php?mission=hst&id=16734}{16734} was provided through a grant from the STScI under NASA contract NAS5-26555. We acknowledge the support from the LBT-Italian Coordination Facility for the execution of observations, data distribution and reduction.
The LBT is an international collaboration among institutions in the United States, Italy and Germany. LBT Corporation partners are: The University of Arizona on behalf of the Arizona university system; Istituto Nazionale di Astrofisica, Italy; LBT Beteiligungsgesellschaft, Germany, representing the Max-Planck Society, the Astrophysical Institute Potsdam, and Heidelberg University; The Ohio State University, and The Research Corporation, on behalf of The University of Notre Dame, University of Minnesota and University of Virginia.

\end{acknowledgments}

%






\appendix

\section{Lensing models}
\label{appendix:lensing}

\cite{Shu+2016b} presented lens models for the seven systems studied here. In order to obtain uncertainties on the sources magnifications we carry out our own modeling.

Our lens mass models are very similar to those of \cite{Shu+2016b}, but the lens reconstructions method is somewhat different. Multiple image lens reconstruction of extended sources, like LAEs, relies on the fact that surface brightness is preserved under lensing mapping from the source to the lens plane. First, from the original F814W HST images we remove all pixels with counts below 0.015/s, because most of these appear to be noise. The remaining pixels that are clearly not a part of the lensed images are also removed.

Our lens mass models consist of a single elliptical {\tt alphapot} potential \citep{Keeton01} with a small core ($< 0.1''$),  $r_c$, centered at $(x_0,y_0)$. The corresponding density profile is a power law of slope $\alpha$. We also include an external shear, whose two parameters, $\gamma_x$ and $\gamma_y$ subsume the influence of outside mass as well as azimuthal complexities in the lens itself. The priors on the alphapot parameters are: the slope of the density profile was $\alpha\in[-0.85,1.15]$, position angle of the lens was limited to $\pm 10^\circ$ of the observed light PA, and the ellipticity of the potential to $b/a>0.8$, corresponding to the ellipticity of the mass distribution of $b/a>0.55$. The center of the lens was allowed to vary within 3 pixels, or $\pm 0.12''$ from the light center of the lensing galaxy.

We backproject lens plane pixels to the source plane, and optimize the parameters of the lens using downhill simplex for 9 free parameters: normalization of the potential, $\alpha$, $r_c$, $x_0$, $y_0$, $\gamma_x$, $\gamma_y$, PA, and $b/a$. Our figure of merit to be minimized is a product of two criteria, both defined in the source plane: (i) since the images that backproject to the same source plane location must arise from the same portion of the source, the rms of fluxes of overlapping backprojected pixels at any source plane location should be small, and (ii) the source plane area occupied by high surface brightness pixels in the lens plane should be small. The second criterion is similar to that used in point-image reconstruction, where point images would ideally converge to the same source plane location. In practice, this can lead to overfocusing and higher implied magnifications, however, in our case this is mitigated by the first criterion. Furthermore, because of the extended nature of the images and their uneven shape, and the restrictions imposed by what lensing deflection fields can do, overfocusing is unlikely to significantly affect magnification estimates.

While we optimize in the source plane only, we do check that the forward lensed images are in reasonable agreement with the observed images. We generate 100 models for each of the seven systems, and use these to derive uncertainties on source magnification. Our resulting reconstructed sources are non-parametric. Their shape and orientation are similar to those \cite{Shu+2016b}. The resulting density slopes of our models tend to be somewhat shallower than isothermal, extending to $\alpha\sim -0.9$. 

We calculate two magnification values for each source: using 0.018 counts/s, and 0.025 counts/s as the lower limit on the flux. We chose these because at fluxes below 0.018 counts/s noise pixels begin to appear everywhere in the image, while above 0.025 counts/s most of the pixels belong to the lensed images. The former magnifications are lower than the latter because down to 0.018 counts/s, images (and sources) are more extended and so many regions of the source are far from caustics, i.e. lines in the source plane where magnifications are infinite for point sources. We note that magnification estimates depend on whether the source is assumed to be smooth or structured, for example, having very peaked central flux density. In our reconstructions the source plane grid resolution is 0.05 of the total extent of the backprojected pixels in each of the two orthogonal directions. If we assumed a grid $4\times$ finer in $x$ and $y$, our magnification values would be 10 \% - 50 \% larger. 
Our median magnification values, and the corresponding values from \cite{Shu+2016b} are in Table~\ref{tab:properties}. We highlight that the lensing magnification is not achromatic, as continuum and \lya\ photons have different
spatial distributions (the \lya\ photons are certainly extended beyond the arc structures and within the BOSS aperture). Overcoming these problems requires angular resolution maps of the \lya\ emission which we do not possess at this time.

\bibliographystyle{aasjournal}
\bibliography{bibliography1.bib}



\end{document}